\documentclass[a4paper,11pt]{article}

\pdfoutput=1 

\usepackage{fullpage}

\usepackage{mathrsfs}
\usepackage{comment}
\usepackage{centernot}
\usepackage{mathtools}
\usepackage{stmaryrd}

\usepackage{latexsym,amssymb,amstext,amsmath}
\usepackage{hyperref}
\usepackage{graphbox,graphicx}
\usepackage{subcaption}
\usepackage{amsthm}
\usepackage{esint}
\usepackage{float}

\def\be {\begin{equation}}
\def\ee {\end{equation}}
\def\bea {\begin{eqnarray}}
\def\eea {\end{eqnarray}}

\def\beq{\begin{equation}}
\def\eeq{\end{equation}}
\def\beqa{\begin{eqnarray}}
\def\eeqa{\end{eqnarray}}

\newtheorem{theorem}{Theorem}[section]
\newtheorem{lemma}[theorem]{Lemma}

\newtheorem{corollary}[theorem]{Corollary}

\theoremstyle{definition}

\newtheorem{remark}[theorem]{Remark}
\newtheorem{definition}[theorem]{Definition}

\begin{document}

\title{\textbf{\LARGE  Riemann-Hilbert problems, Toeplitz operators \\ and ergosurfaces}}
\author{M.~Cristina C\^amara \ and \ Gabriel Lopes Cardoso}
\date{\small 
\vspace{-5ex}
\begin{quote}
\emph{
\begin{itemize}
\item[]
Center for Mathematical Analysis, Geometry and Dynamical Systems,\\
Department of Mathematics, 
  Instituto Superior T\'ecnico, Universidade de Lisboa,\\
 Av. Rovisco Pais, 1049-001 Lisboa, Portugal
  \end{itemize}
}
\end{quote}
{\tt cristina.camara@tecnico.ulisboa.pt, gabriel.lopes.cardoso@tecnico.ulisboa.pt}
}

\maketitle
%


\begin{abstract}
\noindent
The Riemann-Hilbert approach, in conjunction with the canonical Wiener-Hopf factorisation of certain matrix functions called monodromy matrices,
enables one to obtain
explicit solutions to the non-linear field equations of some gravitational theories. 
These solutions are encoded in the elements of a matrix $M$ depending on the Weyl coordinates $\rho$ and $v$, determined
by that factorisation.
We address here, for the first time,
 the underlying question of what happens when a canonical Wiener-Hopf factorisation does not exist, using
the close connection of Wiener-Hopf
factorisation with Toeplitz operators to study this question.
For the case of rational monodromy matrices, we prove that the non-existence of a canonical Wiener-Hopf factorisation determines curves
in the $(\rho,v)$ plane on which some elements of $M(\rho,v)$ tend to infinity, but 
where the space-time metric may still be well behaved.
In the case of uncharged 
rotating black holes in four space-time dimensions and, for certain choices of coordinates, in five space-time dimensions, we show that these curves correspond to their ergosurfaces.

\end{abstract}

\vskip 5mm


\section{Introduction  }

It is a remarkable and rather surprising fact that a great variety of problems in mathematics, physics and engineering   -- in diffraction theory, elastodynamics, control theory, 
 integrable systems 
and, more recently, in the study
of various compressions of multiplication operators such as truncated and dual-band Toeplitz operators  -- can be reformulated as, or reduced to
a Riemann-Hilbert problem \cite{Du,MS,Gohberg2003FactorizationAI,Its,CP,De,COP,KisilAMR,KA}.

One  field of application of the Riemann-Hilbert approach is the study of the solution space of the Einstein field equations in $D$ space-time dimensions. Building on the work of \cite{Breitenlohner:1986um,Lu:2007jc},
this field of application has become the object of revived interest, 
see for instance \cite{Katsimpouri:2012ky,Chakrabarty:2014ora,Camara:2017hez,Cardoso:2017cgi,Aniceto:2019rhg, Penna:2021kua, Camara:2022gvc}
and references therein. Here we study the cases of $D=4$ and $D=5$.
The resulting PDE's, after reduction to two dimensions, form an integrable system \cite{Breitenlohner:1986um,Lu:2007jc}, i.e., they appear as a compatibility condition for an auxiliary linear system, called a Lax pair 
\cite{Its}. The Lax pair that underlies the work of \cite{Breitenlohner:1986um,Lu:2007jc} is called 
Breitenlohner-Maison linear system. There exists another Lax pair, called 
Belinski-Zakharov linear system, which has been shown to be equivalent to the Breitenlohner-Maison linear system \cite{Figueras:2009mc}, and which is particularly well suited to obtain exact solutions of solitonic type from a seed solution by the inverse scattering method \cite{Belinsky:1971nt}. Even though these two linear systems are equivalent, they originate different approaches to solving the field equations. The approach based on the Breitenlohner-Maison linear
system, which we will follow here, does not require knowledge of a seed solution and has the advantage of 
making use of the group structure that underlies the dimensionally reduced model.

The Breitenlohner-Maison linear system \cite{Breitenlohner:1986um,Lu:2007jc} is a Lax pair depending on two real coordinates
which will be denoted by $(\rho, v)$ (Weyl coordinates). It also
depends on a complex parameter $\tau$ that varies 
on an algebraic curve, called the spectral curve, that depends on $(\rho, v)$. 
The Riemann-Hilbert approach based on this Lax pair
allows for the explicit construction of solutions of the Einstein field equations by means of the canonical 
Wiener-Hopf factorisation of monodromy matrices with respect
to a contour $\Gamma$ in the complex $\tau$-plane, as proven in \cite{Aniceto:2019rhg}.
In this factorisation, the coordinates $(\rho, v)$ play the role of parameters.

This approach necessarily assumes the existence of a canonical Wiener-Hopf factorisation. But does the latter always exist? And what happens
if it does not?

Here we use for the first time the close connection of Toeplitz operators with Wiener-Hopf factorisation to establish necessary and sufficient
conditions for the factorisation of a rational monodromy matrix to be canonical, and we show that a canonical Wiener-Hopf factorisation may not exist on certain curves
$D(\rho, v) =0$.

Note that, due to the role of the spectral curve in the case that we are studying, the question addressed here is different from the well studied question
in factorisation theory, of when does a general rational matrix in the variable $\tau$ have a canonical Wiener-Hopf factorisation. For instance, it follows from  a theorem
in \cite{Cardoso:2017cgi} that {\it any} monodromy matrix which can be reduced to triangular form by multiplication by constant matrices
admits a canonical Wiener-Hopf factorisation, in sharp contrast with what happens with general triangular matrices 
\cite{CG,MP,LS,AAM,GKR}.

When performing the canonical Wiener-Hopf factorisation of a monodromy matrix ${\cal M}_{\rho, v} (\tau)$  with respect to a certain contour $\Gamma$, one of the factors that arise in the factorisation determines the space-time solution. This is the factor 
${\cal M}^-_{\rho, v} (\tau)$,
see \eqref{defM}. The limit of ${\cal M}^-_{\rho, v} (\tau)$ 
when $\tau \rightarrow \infty$, $M(\rho,v)$, encodes the space-time metric,
which can be obtained from the elements of $M(\rho,v)$ as explained 
in \cite{Breitenlohner:1986um,Chakrabarty:2014ora}, see also Section \ref{lsc} and Appendix \ref{sec:A}.

A natural question
then is: what happens to $M(\rho,v)$  when $(\rho,v)$ approaches the points on a curve
$D(\rho, v) =0$ where the monodromy matrix does not admit a canonical factorisation? We show that elements 
in the matrix $M(\rho, v)$ blow up when approaching such points.
Since the matrix $M(\rho, v)$ determines the space-time metric,
we study the behaviour of the latter for $(\rho,v)$ on the curve $\cal C$ defined by $D(\rho,v)=0$. We show that in four dimensions this corresponds to $g_{tt}$, the component in the line element which is proportional to $dt^2$, vanishing and, although some elements of $M(\rho,v)$ may tend to infinity as one
approaches $\cal C$, this does not imply that the space-time metric is ill-behaved on that curve.
In the case of the four-dimensional non-extremal Kerr
black hole in General Relativity it can be verified that the space-time metric is well behaved for all $(\rho,v)$ and that 
the locus $D(\rho, v) =0$ corresponds
to the ergosurface of the black hole, where the norm of the Killing vector $\partial/\partial t$ vanishes. This is shown in Section \ref{sec:kerrMP}, where we also discuss 
the case of the five-dimensional rotating Myers-Perry black hole carrying one angular momentum. 
We show that in this case the space-time metric is also well behaved but now $g_{tt}$ may or may not vanish when $(\rho,v)$ lies on the curve $\cal C$, depending on the choice of coordinates in five dimensions. 

These results moreover shed new light on the relations between the matrix $M(\rho,v)$ and the corresponding space-time metric, showing that a singular behaviour in the elements of $M(\rho,v)$ may not be reflected in the space-time metric.

The paper is organised as follows. To keep it as self-contained as possible, in Section  \ref{lsc}
we briefly review the description of the dimensionally
reduced gravitational field equations as an integrable system associated with a Lax pair (the Breitenlohner-Maison linear system), and we describe
how to construct solutions to the field equations by means of canonical Wiener-Hopf factorisations, if the latter exist.
In Section  \ref{sec:TOWH}, which contains 
the main results of the paper, we use the close relation of canonical Wiener-Hopf factorisation with Toeplitz operators to show that the canonical  factorisation of a rational monodromy matrix may not exist on certain curves in the $(\rho,v)$ plane, and that elements of $M(\rho,v)$ blow up there. We prove this for general $2 \times 2$ monodromy matrices,
but our approach can be extended to the case of $n \times n$ rational monodromy matrices in a straightforward manner. In Section 
\ref{sec:kerrMP}, we illustrate the above with two examples. In the first example we discuss a $2 \times 2$ monodromy matrix, whose
canonical Wiener-Hopf factorisation with respect to a suitably chosen contour $\Gamma$ yields a space-time solution describing the exterior region of the four-dimensional non-extremal Kerr
black hole in General Relativity. We show that in this case the curve $D(\rho, v) =0$ on which the canonical Wiener-Hopf factorisation ceases to exist is the curve for the ergosurface of the black hole.  In the second example, we consider two 
$3 \times 3$ monodromy matrices. In both cases, their
canonical Wiener-Hopf factorisation with respect to suitably chosen contours $\Gamma$ yield the same Myers-Perry solution, although
written in different coordinates. This solution describes the exterior region of a five-dimensional rotating black hole with one angular momentum. In each of these cases, the canonical Wiener-Hopf factorisation ceases to exist on a certain curve. It coincides
with the ergosurface of the black hole in one case. In the other case,
there is no such correspondence, and we see that the choice of coordinates in five space-time dimensions allows for a matrix $M(\rho,v)$ which is well behaved on the ergosurface.
Finally, in Appendix \ref{sec:A} we give a few details of the canonical Wiener-Hopf factorisation of one of the $3 \times 3$ monodromy matrices mentioned above.

\section{Solving the field equations by canonical Wiener-Hopf factorisation \label{lsc}}

The field equations of gravitational theories in $D$ space-time dimensions 
form a system of non-linear PDE's for the space-time metric and matter fields.
Due to their non-linear nature, obtaining exact solutions to these PDE's is a highly non-trivial task.
By restricting attention to the subclass of solutions that only depend on two of the $D$  space-time coordinates,
the field equations become effectively two-dimensional, and in certain cases
powerful methods for constructing solutions to these field equations become available. This is the case 
of the gravitational theories discussed in  \cite{Nicolai:1991tt,Schwarz:1995af,Lu:2007jc}, whose two-step reduction to two dimensions
yields the following matricial non-linear field equation in two-dimensions,
\bea 
\label{emotion}
d \left( \rho \star A \right) = 0 \;\;\;,\;\;\; \text{with} \; \; A = M^{-1} dM  \;,
\eea
where $M \in G/H$ is a coset representative of the symmetric space $G/H$ that arises in the two-step reduction \cite{Breitenlohner:1987dg}.
$G/H$ is invariant under an involution $\natural$ called {\it generalized transposition}, i.e.  $M^{\natural} = M $, and $M$
depends on two coordinates, denoted here by $\rho$ and $v$, with $\rho > 0$ and $v \in \mathbb{R}$, called Weyl coordinates.
In \eqref{emotion}, 
$\star$ denotes the Hodge star operator in two dimensions; we have that
\bea
\star d \rho = - \lambda \, dv \;\;\;,\;\;\; \star dv = d \rho \;\;\;,\;\;\; (\star)^2 = - \lambda \, {\rm id} \;,
\label{stho}
\eea
where $\lambda = \pm 1$ depending on whether $(\rho,v)$ are both space-like coordinates ($\lambda =1$) or whether
one of the two is time-like ($\lambda =-1)$. When $D=4$, a solution $M$ to \eqref{emotion} yields a gravitational solution whose four-dimensional space-time metric 
is given by
\bea
ds_4^2 = - \lambda \, \Delta (dt  + B  d \phi  )^2 + \Delta^{-1} 
\left(  e^\psi \, ds_2^2 + \rho^2 d\phi^2 \right) \;,
\label{4dWLP}
\eea
where $ ds_2^2 = \sigma d\rho^2 + \varepsilon dv^2$ with 
$\sigma \varepsilon = \lambda$, 
$\Delta$ and $B$ are functions of $(\rho, v)$ determined  by the solution $M(\rho,v)$ of \eqref{emotion} and
$\psi(\rho,v)$ is a scalar function
determined  from  $M(\rho,v)$  by integration \cite{Schwarz:1995af,Lu:2007jc}, see also eq. (2.7) in \cite{Aniceto:2019rhg}.
As an example, consider the two-step reduction of the four-dimensional Einstein-Hilbert action to two dimensions.
The resulting coset is $G/H = SL(2, \mathbb{R})/SO(2)$, the involution $\natural$ is matrix transposition and the coset representative $M$ takes the form 
\bea
\label{M22DB}
M =
\begin{pmatrix}
 \Delta + {\tilde B}^2/\Delta &\;  {\tilde B}/ \Delta\\
{\tilde B}/\Delta  &\;  1/\Delta
\end{pmatrix} \;,
\eea
where ${\tilde B}$ is related with $B$ through $ \rho \star d {\tilde B} = \Delta^2 \, dB $
\cite{Breitenlohner:1986um}.

The non-linear field equation \eqref{emotion} is the compatibility condition for an auxiliary
linear system of differential equations (a Lax pair) given by 
\begin{equation}
	\tau \left( dX + A X \right) = \star \, dX \;
	\label{laxBM}
\end{equation}
(\cite{Lu:2007jc}) called the Breitenlohner-Maison linear system \cite{Breitenlohner:1986um}.
Here $\tau$ denotes a complex parameter, called spectral parameter, associated through the following algebraic relation with another complex variable $\omega$ and the coordinates $(\rho,v)$,
\bea
\omega = v + \frac{\lambda}{2}    \, \rho \, \frac{\lambda -   \tau^2}{\tau} \;\;\;,\;\;\; \tau \in \mathbb{C} \backslash \{0\} \;,
\label{spect}
\eea
which we will refer to as the spectral relation.
The significance of the linear system \eqref{laxBM} is explained in \cite{Lu:2007jc,Aniceto:2019rhg}.
One powerful method to obtain a pair $(X,A)$ where  $X$ is a solution 
to the Lax pair \eqref{laxBM} with input  $A = M^{-1} d M$, and $M$ is a solution to the field equation \eqref{emotion}, consists in obtaining 
a canonical bounded Wiener-Hopf (WH) factorisation, also known as Birkhoff factorisation, of
a so-called monodromy matrix (\cite{Aniceto:2019rhg}), defined as follows.

Let ${\cal M} (\omega)$ be a $\natural$-invariant $n \times n$ matrix function of the complex variable $\omega$ and denote by ${\cal M}_{\rho, v} (\tau)$ the matrix
that is obtained from  ${\cal M} (\omega)$ by composition using the spectral relation \eqref{spect}, i.e.,
\bea
{\cal M}_{\rho, v} (\tau) = {\cal M} \left( v + \frac{\lambda}{2}    \, \rho \, \frac{\lambda -   \tau^2}{\tau}  \right) \,.
\label{mont} 
\eea
We call ${\cal M}_{\rho, v} (\tau)$ a monodromy matrix.

Let now
$\Gamma$ be a simple closed contour in $\mathbb{C}$, encircling the origin, and invariant under the involution
 $\tau \mapsto
- \lambda/\tau $, and let
 $\mathbb{D}^+_{\Gamma}$ and $\mathbb{D}^-_{\Gamma}$ denote the interior and the exterior of $\Gamma$  (including the point $\infty$), respectively.

 {\definition
A bounded Wiener-Hopf (WH) factorisation 
of ${\cal M}_{\rho, v} (\tau)$ with respect to $\Gamma$ is a decomposition 
of the form 
 \bea
{\cal M}_{\rho, v} (\tau)= {\cal M}^-_{\rho, v} (\tau)  \, D(\tau)  \, {\cal M}^+_{\rho, v} (\tau)  \;\;\;,\;\;\; 
\tau 
\in \Gamma 
\label{MMDM}
\eea
such that the $n \times n$ matrix functions $ {\cal M}^{\pm}_{\rho, v} (\tau)$, as well as their inverses, 
admit an 
 analytic and bounded extension to $\mathbb{D}^{\pm}_{\Gamma}$, respectively, and $D(\tau)$ is a diagonal matrix with
 diagonal elements of the form $\tau^{k_i}$, $i = 1, 2, \dots, n$, with $k_i \in \mathbb{Z}$. If $D(\tau) = \mathbb{I}_{n \times n}$, i.e. $k_i = 0$ for all $i=1, 2, \dots, n$, then the factorisation 
 \bea
{\cal M}_{\rho, v} (\tau)= {\cal M}^-_{\rho, v} (\tau)  \, {\cal M}^+_{\rho, v} (\tau)  \;\;\;,\;\;\; 
\tau 
\in \Gamma 
\label{MMM}
\eea
is said to be {\it canonical}.
}\\

We will use the abbreviated forms WH factorisation and canonical WH factorisation to denote \eqref{MMDM} and \eqref{MMM}, respectively.

We will consider here only matrix functions which, for fixed $(\rho,v)$ and variable $\tau$, have elements in the algebra $C^{\mu} (\Gamma)$ of all H\"older continuous functions on $\Gamma$ with exponent $\mu \in \,] 0, 1[$ (\cite{MP}). It is well known that every matrix function in $(C^{\mu} (\Gamma))^{n \times n}$, invertible in this algebra, admits a WH factorisation with factors in the same algebra. If this factorisation is canonical, then it is unique once one imposes the normalisation condition 
\bea
{\cal M}^+_{\rho, v} (0) = \mathbb{I}_{n \times n} \;.
\eea
With this normalisation condition, we denote
\bea
{\cal M}^+_{\rho, v} (\tau) =: X(\tau, \rho, v) \;.
\label{MpX}
\eea

It was shown in Theorem 6.1 of \cite{Aniceto:2019rhg} that, under 
very general assumptions, if a 
canonical WH factorisation of ${\cal M}_{\rho, v} (\tau)$ with respect to a contour $\Gamma$ 
(satisfying the conditions above) exists, then a 
solution $M$ to the field equation \eqref{emotion} is given by
\bea 
M(\rho,v) : = \displaystyle{\lim_{\tau \rightarrow \infty}} {\cal M}^-_{\rho, v} (\tau) ,
\label{defM}
\eea
while $X(\tau, \rho,v)$ is a solution to the linear system  \eqref{laxBM} with input $A = M^{-1} d M$.

Two questions naturally arise at this point. The first is the question of {\it existence} of a canonical WH factorisation. The second is the question of the {\it behaviour} of the solution $M(\rho,v)$, obtained from a canonical WH factorisation, when $(\rho,v)$ approaches a point $(\rho_0, v_0)$ in the plane of the Weyl coordinates for which a canonical WH factorisation of ${\cal M}_{\rho_0, v_0} (\tau)$ does not exist.
We address these two questions in the next section.

Note that, since every invertible matrix in $(C^{\mu} (\Gamma))^{n \times n}$,
in the variable $\tau$, with determinant equal to $1$ (as will be our case), has a WH factorisation (\cite{MP}), for such a point $(\rho_0, v_0)$ the corresponding monodromy matrix will have a {\it non canonical} WH factorisation \eqref{MMDM}. However, there is no point in determining the latter if one wants to answer the question of behaviour of the solutions, since the factors \eqref{MMDM}
are not the limit, when $(\rho, v) \rightarrow (\rho_0, v_0)$, of those in a canonical factorisation valid for points $(\rho,v)$ in a neighbourhood of $(\rho_0,v_0)$.

\section{Toeplitz operators, Wiener-Hopf factorisation and ergosurfaces \label{sec:TOWH}}

To address the fundamental question of {\it existence} of a canonical WH factorisation for a monodromy matrix underlying all results obtained in the literature using the Breitenlohner-Maison factorisation approach, we will use the close connection of WH factorisation with Toeplitz operators. We start by briefly explaining this relation.

Let $\Gamma$ be a simple closed
curve $\Gamma$ in the complex $\tau$-plane encircling $0$, positively oriented and invariant under the involution $\tau \mapsto -\lambda/\tau$.
We call such a contour {\it an admissible contour}.
Let us 
denote by $S_{\Gamma}$ the singular integral operator with Cauchy kernel on $L^2 (\Gamma)$,
\bea
\left( S_{\Gamma} f \right) (\tau) = \frac{1}{\pi i} \, {\rm p.v.} \int_{\Gamma} \, \frac{f(z)}{z - \tau} \, dz \;\;\; , \;\;\; \tau \in \Gamma, 
\eea
where $\rm p.v.$ denotes Cauchy's principal value. Then one can define two complementary orthogonal projections 
\bea
P^{\pm}_{\Gamma} = \frac12 \left( I \pm S_{\Gamma} \right) \;.
\eea
The functions in
$H^2_{+} := P^{+}_{\Gamma} L^2 (\Gamma)$ have an analytic extension to the interior of $\Gamma$, while
the functions in $H^2_{-} := P^{-}_{\Gamma} L^2 (\Gamma)$ have an analytic extension to the exterior of $\Gamma$ and vanish at $\infty$. We can write
\bea
L^2 (\Gamma) = H^2_{-} \oplus H^2_{+} \,,
\eea
so every function in $L^2(\Gamma)$ admits a unique decomposition $f= f_-+f_+$ with $f_{\pm} \in H^2_{\pm}$.
It may be shown that $S_{\Gamma}$ maps $C^{\mu}(\Gamma)$ into  $C^{\mu}(\Gamma)$ and 
$C_+^{\mu}(\Gamma) := P^{+}_{\Gamma} C^{\mu}(\Gamma) $, $C_-^{\mu}(\Gamma):= 
P^{-}_{\Gamma} C^{\mu}(\Gamma) \oplus \mathbb{C} $ are closed subalgebras of $C^{\mu} (\Gamma)$. For this reason, $C^{\mu}(\Gamma)$ is called a decomposing algebra (\cite{MP}). Every invertible element of $C^{\mu}(\Gamma)$ has a WH factorisation with factors in the same algebra, and this result extends to matrix functions in 
$(C^{\mu} (\Gamma))^{n \times n}$ (\cite{MP}).  If an $n \times n$ matrix $G \in (C^{\mu} (\Gamma))^{n \times n}$
admits a canonical WH factorisation of the form
\bea
G = G_- G_+ \;,
\label{ggmgp}
\eea
then the factors must be in the same algebra, with
\bea
G^{\pm 1}_{-} \in  (C_-^{\mu} (\Gamma))^{n \times n} \;\;\;,\;\;\; G^{\pm 1}_{+} \in  (C_+^{\mu} (\Gamma))^{n \times n} \;.
\label{gcmgcp}
\eea
The factorisation is unique if we impose moreover the normalising condition 
\bea
G_+(0) =  \mathbb{I}_{n \times n} \;.
\label{Gnorm}
\eea
However, the question of determining whether or not the factorisation is canonical is a rather nontrivial one in general for matrix functions. It turns out that this question may be formulated in terms of Toeplitz operators.

Toeplitz operators are compressions of multiplication operators into the Hardy space $H^2_{+}$ (or its vectorial analogue, in the matricial case). Concretely, given an $n \times n$ matrix function $G$ whose elements are bounded functions on $\Gamma$, the Toeplitz operator $T_G$ is defined by
\bea
T_G = P^+ G P^+\vert_{(H^2_+)^n} : (H^2_+)^n \rightarrow (H^2_+)^n .
\eea
$G$ is called the symbol of the operator $T_G$. There exists a close connection between the study of Toeplitz operators and the theory of WH factorisation. Indeed, the operator $T_G$ is Fredholm, i.e. it has a closed range and finite dimensional kernel and cokernel, if and only if $G$ admits a WH factorisation; $T_G$ is invertible if and only if that factorisation is canonical (see \cite{CDR}) and references therein).

It follows from the above results on WH factorisation of matrix functions in $(C^{\mu} (\Gamma))^{n \times n}$ that, if $G$ is invertible in that algebra, $T_G$ is always Fredholm 
\cite{CG, MP, LS}. In that case, assuming moreover
that $\det G=1$, we have that
\bea
\dim \ker T_G - \text{codim  Im}  T_G = 0 ,
\label{kTcodT}
\eea
i.e., the Fredholm index of $T_G$ is zero. 
Here $\ker T_G$ denotes the kernel of $T_G$, Im $T_G$ denotes its range and codim Im $T_G = 
 \dim \left( (H^2_+)^n / \text{Im} T_G \right) $.
It follows from \eqref{kTcodT} that 
$T_G$ is invertible if and only if it is
injective. As a consequence, $G$ has a canonical WH factorisation if and only if $\ker T_G = \{ 0 \}$, or,
equivalently, the Riemann-Hilbert problem
\bea
G \phi_+ = \phi_- \quad , \quad \text{with}  \quad \phi_{\pm} \in (H^2_{\pm})^n ,
\label{rhgH}
\eea
whose solutions $\phi_+$ constitute the kernel of $T_G$, admits only the zero solution. We summarize the above relations as follows, 
 taking \eqref{ggmgp} - \eqref{Gnorm} into account.

{\theorem \label{theo81}
If $\Gamma $ is an admissible contour in the complex plane and 
$G \in (C^{\mu}(\Gamma))^{n \times n}$, $\mu \in \, ]0,1[$, with $\det G = 1$, 
then $G$ has a canonical WH
factorisation if and only if the Toeplitz operator $T_G$ in $(H^2_{+})^n$
is injective, i.e., if and only if
\eqref{rhgH} admits only the trivial solution $\phi_{\pm} =0$. In that case
the factors $G_{\pm}$ in a canonical WH factorisation
$G = G_- G_+$ belong to $(C_{\pm}^{\mu} (\Gamma) )^{n \times n}$, as well as their inverses.
The $\rm i^{th} $ columns of $G_+^{-1}$ and $G_-$ are given by the unique solution
of the Riemann-Hilbert problem 
\bea
G \psi_+^i = \psi_-^i \quad \text{with} \quad  \psi_{\pm}^i \in (C^{\mu}_{\pm} (\Gamma))^n 
\label{gpsps}
\eea
satisfying the normalising condition
$\psi^i_+(0) = [ \delta_{ij} ]^T_{j=1, \dots, n}$. 
}\\

{\remark \label{remgm}

Note that, although formally similar, \eqref{rhgH} and \eqref{gpsps} are different Riemann-Hilbert problems, since we seek their solutions
in different spaces, given that they address different questions. By solving \eqref{rhgH}, which determines the kernel of $T_G$ in $(H^2_+)^n$, we answer the question of whether or not there exists a canonical WH factorisation. In particular, since 
we must have $\phi_{-} \in (H^2_-)^n $, we impose that $\phi_- (\infty) = 0$. If a canonical WH factorisation of $G$ does exist, i.e., if \eqref{rhgH} admits only the zero solution, then \eqref{gpsps} necessarily has $n$ linearly independent solutions $(\psi_+, \psi_-)$, with 
$\psi_{\pm} \in (C^{\mu}_{\pm} (\Gamma))^n  $. Note that in this case we only impose that $\psi_-$ must be bounded at $\infty$.
The solutions to \eqref{gpsps} provide the columns of the factors $G_+^{-1}$ and $G_-$.}
\\

We now apply these results to monodromy matrices obtained from
rational matrix functions ${\cal M} (\omega)$ of the form
\bea
{\cal M} (\omega) = \frac{1}{q (\omega) }  \begin{pmatrix}
p_{11} (\omega) & p_{12} (\omega)  \\
p_{12} (\omega) & p_{22} (\omega)
\end{pmatrix},
\label{pomeg}
\eea
where $q, p_{11}, p_{12}, p_{22}$ denote polynomials of degree $n, k_{11}, k_{12}, k_{22}$, respectively, 
and $\det {\cal M} =1 $, i.e., 
\bea
q^2 = p_{11} p_{22} - (p_{12})^2 \,.
\label{qprel}
\eea
Note that ${\cal M}$ must be a symmetric matrix, since $\natural$ is matrix transposition in this case and we must have ${\cal M} = {\cal M}^{\natural}$. For simplicity, 
and taking the applications into account, we also assume that the zeroes of $q$ are all of order $1$
and $\lambda =1$.

{\remark

It is easy to see that the approach presented in this section to address the questions formulated at the end of Section \ref{lsc} can be extended to the case of $n \times n$ rational matrix functions (examples will be given in Section \ref{sec:5drot} for $3 \times 3$ monodromy matrices)
and does not depend on the order of the zeroes of $q$ nor on
the particular value of $\lambda$.
}\\

Let us denote the right hand side of \eqref{spect}, with $\lambda =1$, by $\omega_{\tau, \rho, v}$. Any non-trivial polynomial $p(\omega)$ of degree $k$, upon composition with $\omega = \omega_{\tau, \rho, v}$, becomes a product of $k$ factors of the form
\bea
\omega_{\tau, \rho, v} - \omega_0 = \frac{ - \frac{\rho}{2} \, \tau^2 + (v - \omega_0) \, \tau + \frac{\rho}{2} }{\tau} \,.
\label{omom0}
\eea
The numerator of the right hand side of \eqref{omom0} is a polynomial of degree 2 in $\tau$, which does not vanish for $\tau =0$, with simple
zeroes at
\bea
\frac{ v - \omega_0 \pm \sqrt{(v - \omega_0)^2 + \rho^2}}{\rho} \,.
\label{zvom}
\eea
These zeroes appear in pairs, of the form
\bea
\{\tau_0, - \frac{1}{\tau_0} \} \,;
\label{zz0}
\eea
{from} each pair we will choose one of the elements. Suppose that the latter is denoted by $\tau_i$; then the other element, $- 1/\tau_i$, will
be denoted by ${\tilde \tau}_i$. Note that, for any admissible contour $\Gamma$, 
if $\tau_i$ does not belong to $\Gamma$ then neither does ${\tilde \tau}_i$, and we necessarily have one of the points in $\mathbb{D}^+_{\Gamma}$ and the other in $\mathbb{D}^-_{\Gamma}$ (\cite{Aniceto:2019rhg}).

With this in mind, for any polynomial $p_k (\omega)$ of degree $k$, let us use the representation (upon composition with  $\omega_{\tau, \rho, v}$)
\bea
p_k ( \omega_{\tau, \rho, v}) = \frac{p_{2k} (\tau)}{\tau^k},
\eea
where we omit the dependence of the numerator on $(\rho,v)$. With this notation, the monodromy matrix associated with 
\eqref{pomeg} takes the form
\bea
{\cal M}_{\rho,v} (\tau) = \frac{\tau^n}{q_{2n} (\tau) } \,  \tilde{\cal M}_{\rho,v} (\tau) 
\;\;\;\text{with} \;\;\;   
\tilde{\cal M}_{\rho,v} (\tau) =  \begin{pmatrix}
\frac{{\tilde p}_{2 k_{11}} (\tau)}{\tau^{k_{11}}}  & \frac{{\tilde p}_{2 k_{12}} (\tau)}{\tau^{k_{12}} }  \\ 
\frac{{\tilde p}_{2k_{12}} (\tau)}{\tau^{k_{12}} } & \frac{{\tilde p}_{2 k_{22}} (\tau)}{\tau^{k_{22}}}
\end{pmatrix} ,
\label{calm22}
\eea
with
\bea
\det {\cal M}_{\rho,v} (\tau) = 
\frac{{\tilde p}_{2 k_{11}} (\tau) \, {\tilde p}_{2 k_{22}} (\tau)}{\tau^{k_{11} + k_{22}}} - \frac{{\tilde p}^2_{2 k_{12}} (\tau)}{\tau^{2 k_{12}}} = \frac{q^2_{2n} (\tau)}{\tau^{2n}}.
\label{q2n}
\eea
Note that, since $q_{2n} (0) \neq 0$ (cf. \eqref{omom0}), we must have, by \eqref{qprel},
\bea
2n = \max \{ k_{11} + k_{22}, 2 k_{12} \} .
\label{2nkk}
\eea
Let moreover, for the first and second rows of 
 ${\tilde {\cal M}}_{\rho,v} (\tau) $, 
\bea
N_1 = \max \{k_{11}, k_{12} \} \;\;\;,\;\;\; N_2 = \max \{k_{12}, k_{22} \} .
\label{NNkk}
\eea

{\remark
\label{remNNk}
It is not difficult to see that the conditions \eqref{2nkk} and \eqref{NNkk} together imply that we cannot have simultaneously $N_1 = k_{11} > k_{12} $ and
$N_2 = k_{12} > k_{22}$, nor can we have  simultaneously $N_1 = k_{12} > k_{11} $ and
$N_2 = k_{22} > k_{12}$.
}\\

In order to formulate our main results in this section, we still need to define the contour with respect to which the factorisation of 
${{\cal M}}_{\rho,v} (\tau) $ is considered.  So, from each of the $n$ pairs of zeroes of $q_{2n} (\tau)$, of the form \eqref{zvom}, we choose one point which we denote by $\tau_i$ ($i=1, \dots, n$),
and we denote the other by ${\tilde \tau}_i$, and we take $\Gamma$ to be any admissible contour such that $\{\tau_i, i= 1, \dots, n \}$ is contained in $\mathbb{D}^+_{\Gamma}$.
Note that
$\Gamma$ may depend on $(\rho, v)$, but we omit this for simplicity of notation, unless necessary.
We assume $(\rho,v)$ to be such that $\rho >0$ and $v \pm i \rho$ are not zeroes of $q$.

We can now state our main theorems. The proofs will be given in Section \ref{sec:proo-theo}.

We start by addressing {\it the question of existence} of a canonical WH factorisation of 
${\cal M}_{\rho,v} (\tau)$ by reducing it to the question of whether the injectivity Riemann-Hilbert problem 
for the Toeplitz operator with symbol $ {\cal M}_{\rho,v} (\tau)$, \eqref{rhgH}, admits only the zero solution.

{
\theorem
\label{theoNNn}
With the notation above, for ${\cal M} (\omega)$ of the form \eqref{pomeg} and  ${\cal M}_{\rho,v} (\tau)$
given by \eqref{mont},
we have that:
\\
(i) if $N_1 + N_2 < 2n$, then ${\cal M}_{\rho,v} (\tau)$ has a canonical WH factorisation w.r.t. $\Gamma$, for all $\rho,v$;\\
(ii) if $N_1 + N_2 = 2n$, then ${\cal M}_{\rho,v} (\tau)$ has a canonical WH factorisation if and only if \bea
D (\rho, v) \neq 0 \,,
\label{Drv}
\eea
where $D(\rho, v)$ is the
determinant of the matrix coefficient of the following linear system of $2n$ equations for $2n$ unknowns,
\bea
\left( Q_{N_1-1} \, \tau^{N_2 - k_{22}} \, {\tilde p}_{2 k_{22}}   - Q_{N_2 -1} \, \tau^{N_1 - k_{12}} \, {\tilde p}_{2k_{12}}  \right) (\tau_i) &=& 0,  \nonumber\\
\left(  Q_{N_1-1} \, \tau^{N_2 - k_{22}} \, {\tilde p}_{2 k_{22}}   - Q_{N_2 -1} \, \tau^{N_1 - k_{12}} \, {\tilde p}_{2k_{12}}  \right)' (\tau_i)  &=& 0 \;\;\;,\;\;\; i=1, 2, \dots, n,
\label{qqN}
\eea
where $Q_{N_1-1}$ and $Q_{N_2-1}$ are unknown polynomials of degree at most $N_1 -1$ and $N_2 -1$, respectively, their coefficients being the $2n = N_1 + N_2$ unknowns of the system;\\
(iii) the case where $N_1 + N_2 > 2 n$ can be reduced to (ii).
}\\

It is easy to see from \eqref{calm22} that,
since $\frac{\tau^n}{q_{2n} (\tau) }$ is a scalar function that, by a theorem given in  
\cite[Section 2]{Cardoso:2017cgi}, 
admits a canonical WH factorisation with respect
to any admissible contour $\Gamma$ where $q_{2n}$ does not vanish,
and that this factorisation can be obtained straightforwardly, the existence
of a canonical WH factorisation for 
${{\cal M}}_{\rho,v} (\tau) $ is equivalent to that of $ {\tilde {\cal M}}_{\rho,v} (\tau) $. 
The latter can be obtained, if it exists, as described in Theorem  \ref{theo81}.
For the particular case $(ii)$ of Theorem \ref{theoNNn}, writing ${\tilde {\cal M}}_{\rho,v} (\tau) =
{\tilde {\cal M}}^-_{\rho,v} (\tau) {\tilde X} (\tau, \rho, v)$ according to 
\eqref{MMM} and \eqref{MpX}, we have the following.

{
\theorem \label{theopp2n}
With the same notation as in Theorem \ref{theoNNn}, if $N_1 + N_2 = 2n$ and $D(\rho, v) \neq 0$, each column of ${\tilde X}^{-1} (\tau, \rho,v)$ is given by 
$\psi_+ = (\psi_{1+}, \psi_{2+})^T$ with
\bea
\psi_{1+} = \frac{ Q_{N_1} \, \tau^{N_2 - k_{22}}  \,   {\tilde p}_{2 k_{22}} - Q_{N_2} \, \tau^{N_1 - k_{12}} \, {\tilde p}_{2 k_{12}} }{q^2_{2n} (\tau) } 
\;\;\;,\;\;\;
\psi_{2+} = \frac{ Q_{N_1}  - \tau^{N_1 - k_{11}} \, \, {\tilde p}_{2 k_{11}}\, \psi_{1+} }{\tau^{N_1-k_{12}} \, {\tilde p}_{2 k_{12}}  } ,
\eea
where the $2n+2$  coefficients of $Q_{N_1}, \, Q_{N_2} $ are uniquely determined by the analyticity of $\psi_{1+} $ and $\psi_{2+}$ in $\mathbb{D}^+_{\Gamma}$, together with
the normalizing conditions $\psi_{1+} (0) = 1,  \psi_{2+} (0) = 0$ for the $1^{\rm st}$ column, and $\psi_{1+} (0) = 0,  \psi_{2+} (0) = 1$ for the second column.
The factor ${\tilde {\cal M}}^-_{\rho,v} (\tau)$ is given by 
${\tilde {\cal M}}^-_{\rho,v} (\tau) = 
{\tilde {\cal M}}_{\rho,v} (\tau) {\tilde X}^{-1} (\tau, \rho, v) $, and both ${\tilde X} (\tau, \rho, v)$ and 
${\tilde {\cal M}}^-_{\rho,v} (\tau)$
are $C^{\infty}$ functions of $(\rho,v)$.
}\\

The equation 
\bea
D(\rho, v) = 0
\label{Drv0}
\eea
defines a curve (or a curve system) $\cal C$ in the plane of the Weyl coordinates $\rho,v$. A natural question that arises from $(ii)$ in Theorem \ref{theoNNn} regards the behaviour of the solution $M(\rho,v)$, obtained as in \eqref{defM}, for points $(\rho,v)$ that {\it do not} satisfy \eqref{Drv0}, as we approach a point $(\rho_0, v_0) \in {\cal C}$. We address this question in the following theorem, where $g_{tt}$ denotes the coefficient of the term proportional to $dt^2$ in \eqref{4dWLP}.

{\theorem
\label{theoD0}
With the same notation as in Theorem \ref{theoNNn},
let $N_1 + N_2 = 2n$ and let $(\rho,v)$ be such that \eqref{Drv} holds, so that ${\cal M}_{\rho,v} (\tau)$ admits a canonical WH factorisation. Let moreover $M(\rho,v)$ be the solution to the field
equation \eqref{emotion} given by \eqref{defM}. If $(\rho_0,v_0)$ is a point on ${\cal C}$ then $g_{tt} = - \lambda \Delta (\rho,v)$ in \eqref{4dWLP} tends to $0$ as $(\rho,v)$ tends to $(\rho_0, v_0)$.
}\\

In the case when the space-time metric \eqref{4dWLP} describes the exterior region of a non-extremal Kerr black hole solution in four space-time dimensions, the vanishing of $g_{tt}$ defines the ergosurface of the rotating black hole.

The previous results can be reformulated in terms of Toeplitz operators as follows.

{\corollary

Let $\Gamma$ be an admissible contour and let ${\cal M}_{\rho,v} (\tau)$ be of the form \eqref{calm22}. If $N_1 + N_2 < 2n$, or $N_1 + N_2 = 2n$ with $D(\rho,v) \neq 0$ for all $(\rho,v)$, then the Toeplitz operator with symbol ${\cal M}_{\rho,v} (\tau)$ is invertible and the space-time metric \eqref{4dWLP}, corresponding to $M(\rho,v)$
defined by \eqref{defM}, is well behaved, for all $(\rho,v)$.

If $N_1 + N_2 = 2n$ and there are points in the Weyl half-plane satisfying \eqref{Drv0}, then the Toeplitz operator with symbol 
${\cal M}_{\rho,v} (\tau)$ is not invertible for $(\rho,v)$ on the curve  $D(\rho,v) = 0$ and, in the space-time metric \eqref{4dWLP}, we have $g_{tt} = 0$ on that curve.

}

\subsection{Proofs
\label{sec:proo-theo}}

The proof of Theorem \ref{theoNNn} is based on various lemmata which we now introduce.

{\lemma
\label{lemmaA}
Let the assumptions of Theorem  \ref{theoNNn} be satisfied and let $\phi_{1+}, \phi_{2+}$ satisfy
 \bea
 \label{ppass}
     \left.
    \begin{array}{lr}
      {p}_{11} (\tau) \, \phi_{1+} +  {p}_{12} (\tau)\, \phi_{2+} &= q_1 (\tau) \\
      {p}_{21} (\tau) \, \phi_{1+} +  {p}_{22} (\tau) \, \phi_{2+}  &= q_2(\tau) 
                  \end{array}\right\} \quad \text{on} \quad \Gamma,
  \eea
where ${p}_{ij}, q_i$ ($i,j = 1,2$) are polynomials such that ${p}_{12}$ and ${p}_{22}$ do not have common zeroes and 
$ {p}_{11} \, { p}_{22} - { p}_{12} \, { p}_{21} \neq 0 $ for all $\tau \in \Gamma$. Then if $\phi_{1+}$ is analytic in $\mathbb{D}^+_{\Gamma}$,
$\phi_{2+}$ is also analytic in $\mathbb{D}^+_{\Gamma}$.

}

\begin{proof}

By Cramer's rule we have
\bea
\phi_{1+} = \frac{ q_1 \, {p}_{22} - q_2 \, {p}_{12}}{{p}_{11} \,  { p}_{22} - {p}_{12} \, {p}_{21}},
\eea
and, from the first and the second equation in \eqref{ppass},
\bea
\label{eqeq}
\frac{q_1 - \, {p}_{11} \, \phi_{1+} }{ {p}_{12} } = \frac{q_2 - \, {p}_{21} \, \phi_{1+} }{ {p}_{22} } = \phi_{2+}.
\eea
Let $\phi_{1+} $ be analytic in $\mathbb{D}^+_{\Gamma}$. Since ${p}_{12}$ and ${p}_{22}$ do not have common zeroes,
in the neighbourhood of any zero of ${p}_{12}$ in the interior of $\Gamma$, the right hand side of \eqref{eqeq} must be analytic, and vice-versa.
Hence, $\phi_{2+} $ be analytic in $\mathbb{D}^+_{\Gamma}$.

\end{proof}

{\lemma
\label{lemmaB}

Let $G$ be an $n \times n$ matrix function and let $T_G$ denote the Toeplitz operator on $(H^2_+)^n$ with symbol $G$.  If all elements of 
$\ker T_G$ vanish at $0$, then $\ker T_G = \{ 0 \}$.

}

\begin{proof}
Let $\phi_+ \in (H^2_+)^n$ be an element of $\ker T_G$, i.e.
\bea
G \phi_+ = \phi_- \quad \text{with} \quad \phi_- \in (H^2_-)^n, 
\eea
and assume that $\phi_+(0) = 0$. Then $\phi_+/\tau \in (H^2_+)^n$, and hence $\phi_+/\tau \in \ker T_G$ because
\bea
G \frac{\phi_+}{\tau} = \frac{\phi_-}{\tau} \in (H^2_-)^n .
\eea
Therefore $\phi_+/\tau$ also vanishes at $0$ and, by repeating the same argument, also $\phi_+/\tau^k \in (H^2_+)^n$ for all $k \in \mathbb{N}$, which
is only possible if $\phi_+ = 0$.

\end{proof}

{\lemma
\label{lemmaD}
Let the assumptions and notation of Theorem \ref{theoNNn} hold.  Then $N_1 + N_2 > 2n$ if and only if 
\bea
k_{11} > k_{12} > k_{22}  \; \lor \; k_{22} > k_{12} > k_{11} .
\label{kkk}
\eea

}

\begin{proof}
Since $2n = \max \{ k_{11} + k_{22}, 2 k_{12} \} $, it is clear that \eqref{kkk} implies $N_1 + N_2 > 2n$. Now let us prove the converse, which is equivalent
to proving \
\bea
\sim \left( k_{11} > k_{12} > k_{22}  \; \lor \; k_{22} > k_{12} > k_{11} \right) \, \Rightarrow \, \sim \left(N_1 + N_2 > 2n \right) .
\eea
Let us therefore assume that $ k_{11} > k_{12} > k_{22}$ and $k_{22} > k_{12} > k_{11}$ are both false. 
Falsity of $ k_{11} > k_{12} > k_{22}$ implies that either 
  (i) $k_{11} \leq k_{12}$
or (ii) $k_{12} \leq k_{22}$. Let us first consider the case (i). Then $N_1 + N_2 = k_{12} + \max \{ k_{12}, k_{22} \}$. If $N_2 = \max \{ k_{12}, k_{22} \} = 
k_{12}$, then
$N_1 + N_2 = 2 k_{12} \geq k_{11} + k_{22}$, and hence $2n = 2k_{12} = N_1 + N_2$. If $N_2 =  \max \{ k_{12}, k_{22} \} = k_{22}$, then either $k_{22} >  k_{12}$ or
$k_{22} = k_{12}$. If $k_{22} >  k_{12}$, for the condition $k_{22} > k_{12} > k_{11}$ to be false, we must have $k_{12} \leq k_{11}$, and hence 
$k_{12} = k_{11}$. Then $N_1 + N_2 = k_{11} + k_{22} > 2 k_{12}$ and hence $N_1 + N_2 = 2n$. On the other hand, if $k_{22} = k_{12}$, then
$N_1 + N_2 = 2 k_{12} \geq k_{11} + k_{12}$, and hence  $N_1 + N_2 = 2n$.

Now let consider the case (ii). Then $N_2 = k_{22}$ and $N_1 + N_2 = \max \{ k_{11}, k_{12} \} + k_{22}$.
If $N_1 =  \max \{ k_{11}, k_{12} \} = k_{11}$, then $N_1 + N_2 =  k_{11} + k_{22} \geq 2 k_{12}$, and hence $N_1 + N_2 = 2n$. If
$N_1 =  \max \{ k_{11}, k_{12} \} = k_{12}$, then either $k_{12} >  k_{11}$ or
$k_{11} = k_{12}$. If $k_{12} >  k_{11}$, for the condition $k_{22} > k_{12} > k_{11}$ to be false, we must have $k_{12} \geq k_{22}$, and hence 
$k_{12} = k_{22}$, in which case  $N_1 + N_2 = 2 k_{12}  > k_{11} + k_{22}$ and hence $N_1 + N_2 = 2n$.
On the other hand, if $k_{12} = k_{11}$, then
$N_1 + N_2 = k_{11} + k_{22} \geq 2 k_{12} $, and hence  $N_1 + N_2 = 2n$.

\end{proof}
\vskip 3mm

Using the above lemmata, we proceed with the proofs of Theorems \ref{theoNNn},\ref{theopp2n} and\ref{theoD0}.

\vskip 3mm

\noindent
{\it Proof of Theorem \ref{theoNNn}:}\\
Let us consider the Riemann-Hilbert problem \eqref{rhgH} for the existence of a canonical Wiener-Hopf factorisation for ${\cal M}_{\rho,v} (\tau) = G$. Since
$\tau^n/q_{2n} (\tau)$ admits a canonical Wiener-Hopf factorisation as mentioned before, 
we are reduced to studying \eqref{rhgH} replacing $G$ by 
${\tilde {\cal M}}_{\rho,v}$.
We then have, using \eqref{calm22},
 \bea
     \frac{{\tilde p}_{2k_{11}}}{\tau^{k_{11}}  }\, \phi_{1+} +  \frac{{\tilde p}_{2k_{12}}}{\tau^{k_{12}}} \, \phi_{2+} &= \phi_{1-} , \nonumber\\
      \frac{{\tilde p}_{2k_{12}}}{\tau^{k_{12}}  }\, \phi_{1+} +   \frac{{\tilde p}_{2k_{22}}}{\tau^{k_{22}}  } \, \phi_{2+}  &= \phi_{2-} ,
\eea
where $ \phi_{1 \pm}, \phi_{2\pm} \in H^2_{\pm}$, which is equivalent to 
\bea
   \tau^{N_1-k_{11}}  {\tilde p}_{2k_{11}} \, \phi_{1+} + \tau^{N_1-k_{12}}  {\tilde p}_{2k_{12}} \, \phi_{2+} &= \tau^{N_1} \, \phi_{1-} , \nonumber\\
   \tau^{N_2-k_{12}}  \, {\tilde p}_{2k_{12}} \, \phi_{1+} +  \tau^{N_2-k_{22}}  \, {\tilde p}_{2k_{22}}  \, \phi_{2+}  &= \tau^{N_2} \, \phi_{2-} .
   \label{tNpp}
\eea
Since, for any $N \in \mathbb{N}\cup \{0\}$, $H_+^2 \cap \tau^N H_-^2 = {\cal P}_{N-1}$,
where $ {\cal P}_{N-1}$ is the space of all polynomials of degree at most $N-1$,
we conclude that both sides of 
the equations \eqref{tNpp} are equal to polynomials $Q_{N_1-1} \in {\cal P}_{N_1-1}$ for the first equation and  $Q_{N_2-1} \in {\cal P}_{N_2-1}$
for the second equation. Then
\bea
   \tau^{N_1-k_{11}}  \, {\tilde p}_{2k_{11}}  \, \phi_{1+} + \tau^{N_1-k_{12}}  \, {\tilde p}_{2k_{12}} \, \phi_{2+} &= Q_{N_1 -1}, \nonumber\\
   \tau^{N_2-k_{12}}  \, {\tilde p}_{2k_{12}} \, \phi_{1+} +  \tau^{N_2-k_{22}}  \, {\tilde p}_{2k_{22}}  \, \phi_{2+}  &=  Q_{N_2 -1},
   \label{tNppp}
\eea
and by Cramer's rule
\bea
\phi_{1+} = \frac{
\begin{vmatrix}
Q_{N_1-1} & \tau^{N_1 - k_{12}} \, {\tilde p}_{2k_{12}}  \\
Q_{N_2-1} & \tau^{N_2 - k_{22}} \,  {\tilde p}_{2k_{22}} 
\end{vmatrix}
}{q^2_{2n} (\tau)} \, \tau^{2n - (N_1 + N_2)} \;,\;
\phi_{2+} = \frac{
\begin{vmatrix}
 \tau^{N_1 - k_{11}} \,  {\tilde p}_{2k_{11}}  
& Q_{N_1-1} 
\\
 \tau^{N_2 - k_{12}} \,  {\tilde p}_{2k_{12}}   & Q_{N_2-1} 
\end{vmatrix}
}{q^2_{2n} (\tau)} \, \tau^{2n - (N_1 + N_2)},
\label{cramphi12}
\eea
where we recall that $q^2_{2n}$ is given by \eqref{q2n}.
If $N_1 + N_2 < 2n$ then $\phi_{1+}$ and $\phi_{2+}$ vanish at $0$ and, by Lemma \ref{lemmaB} we must have $\ker T_{{\tilde {\cal M}}_{\rho,v}} = \{0\}$,
and hence ${\tilde {\cal M}}_{\rho,v}$ has a canonical Wiener-Hopf factorisation. If $N_1 + N_2 = 2 n$, then $\phi_{2+}$ will be analytic if  $\phi_{1+}$ is analytic
in $\mathbb{D}^+_{\Gamma}$ by Lemma \ref{lemmaA}, and therefore we are reduced to studying the condition for $\phi_{1+}$ to be analytic.
This corresponds to imposing that the zeroes of the numerator of
\bea
\phi_{1+} = \frac{ Q_{N_1-1} \,  \tau^{N_2 - k_{22}} \,  {\tilde p}_{2k_{22}}  - Q_{N_2-1} \, \tau^{N_1 - k_{12}} \, {\tilde p}_{2k_{12}} }{q^2_{2n} (\tau)}
\eea
cancel the zeroes of $q^2_{2n} (\tau)$ in the interior of $\Gamma$, as in 
\eqref{qqN}. The result of (ii) in Theorem \ref{theoNNn} follows from here.

Finally, if $N_1 + N_2 > 2n$, then by Lemma \ref{lemmaD} we must have either $k_{11} > k_{12} > k_{22}$ or $k_{22} > k_{12} > k_{11}$. 
Let us consider the case $k_{11} > k_{12} > k_{22}$ (the other case can be dealt with in a similar manner).
It follows that $N_1 = k_{11}, N_2 = k_{12}$. Now recall the definition of $n$ given in \eqref{2nkk}.
Let us assume that $2n = k_{11} + k_{22}$ (the other case, $2n = 2 k_{12}$, can be dealt with in a similar manner). Then
\bea
k_{11} + k_{22} = 2n <  N_1 + N_2 = k_{11} + k_{12}.
\eea
Then, \eqref{tNppp} can be written as 
\bea
   {\tilde p}_{2k_{11}} \, \phi_{1+} + \tau^{k_{11}-k_{12}}  {\tilde p}_{2k_{12}} \, \phi_{2+} &= Q_{N_1 -1}, \nonumber\\
    {\tilde p}_{2k_{12}} \, \phi_{1+} +  \tau^{k_{12}- 2n + k_{11}}  \, {\tilde p}_{2k_{22}}  \, \phi_{2+}  &=  Q_{N_2 -1},
 \eea
and by Cramer's rule
\bea
\phi_{1+} &=& \frac{
\begin{vmatrix}
Q_{N_1-1} & \tau^{k_{11} - k_{12}} \,  {\tilde p}_{2k_{12}}\\
Q_{N_2-1} & \tau^{k_{11} - 2n  + k_{12}} \,  {\tilde p}_{2k_{22}}
\end{vmatrix}
}{\tau^{k_{11} - 2n  + k_{12}} \,  {\tilde p}_{2k_{11}} \,  {\tilde p}_{2k_{22}}- \tau^{k_{11} - k_{12}} \,  {\tilde p}_{2k_{12}}^2 } = 
\frac{
\begin{vmatrix}
Q_{N_1-1} & \tau^{2n - 2 k_{12}} \,  {\tilde p}_{2k_{12}} \\
Q_{N_2-1} &    {\tilde p}_{2k_{22}}
\end{vmatrix}
}{{\tilde p}_{2k_{11}} \,  {\tilde p}_{2k_{22}} - \tau^{2n - 2k_{12}} \,  {\tilde p}_{2k_{12}}^2 } \nonumber\\
&=& 
\frac{
\begin{vmatrix}
Q_{N_1-1} & \tau^{2n - 2 k_{12}} \,  {\tilde p}_{2k_{12}} \\
Q_{N_2-1} &    {\tilde p}_{2k_{22}}
\end{vmatrix}
}{q_{2n}^2 (\tau)},
\label{cramp1cram}
\eea
where we used  \eqref{q2n}.
Note that $2n \geq 2 k_{12}$, and hence the powers of $\tau$ in \eqref{cramp1cram} are all non-negative.
Therefore, the analyticity of $\phi_{1+}$ follows by imposing that the numerator in \eqref{cramp1cram} has the same zeroes as $q_{2n}^2 (\tau)$
in the interior of $\Gamma$, i.e. $2n$ zeroes (counting their multiplicity), as it happens when $N_1 + N_2 = 2n$.
\begin{flushright}  \qedsymbol{} \end{flushright}

\vskip 3mm

\noindent
{\it Proof of Theorem \ref{theopp2n}:}\\
Let us consider the Riemann-Hilbert problem \eqref{gpsps} which, in this case, can be written (analogously to the proof of Theorem \ref{theoNNn}) as 
\bea
\tau^{N_1 - k_{11}} \, {\tilde p}_{2 k_{11}} \, \psi_{1 +}  + \tau^{N_1 - {k_{12}}} \, {\tilde p}_{2 k_{12}} \, \psi_{2 +}  = \tau^{N_1} \, \psi_{1-} &=& Q_{N_1} , \nonumber\\
\tau^{N_2 - k_{12}} \, {\tilde p}_{2 k_{12}} \, \psi_{1 +}  + \tau^{N_2 - k_{22}} \, {\tilde p}_{2 k_{22}} \,  \psi_{2 +}  = \tau^{N_2} \, \psi_{2-} &=& Q_{N_2} , 
\label{ppsi}
\eea
where now $ \psi_{i +} \in {C}_+^{\mu} , \,  \psi_{i -} \in {C}_-^{\mu} $ and $Q_{N_1}, \, Q_{N_2} $ are polynomials of degree $N_1, N_2$ (respectively), at most. We have, by Cramer's rule,
\bea
\psi_{1+} = \frac{ Q_{N_1} \, \tau^{N_2 - k_{22}}  \,   {\tilde p}_{2 k_{22}} - Q_{N_2} \, \tau^{N_1 - k_{12}} \, {\tilde p}_{2 k_{12}} }{q^2_{2n} (\tau) },
\label{psi12p}
\eea
and $\psi_{2+}$ is obtained from $\psi_{1+} $ by 
\bea
\psi_{2+} = \frac{ Q_{N_1}  - \tau^{N_1 - k_{11}} \, \, {\tilde p}_{2 k_{11}}\, \psi_{1+} }{\tau^{N_1-k_{12}} \, {\tilde p}_{2 k_{12}}  } ,
\eea
where the $2n + 2$ coefficients of $Q_{N_1} $ and $Q_{N_2}$ are determined by imposing zeroes in the numerator of the right hand side of 
\eqref{psi12p}, so that $\psi_{1+}$ is analytic in the interior of $\Gamma$,
together with the normalization conditions.
Note that the analyticity of 
$\psi_{2+}$ then follows from there according to Lemma \ref{lemmaA}.


Since we assumed that the zeroes of $q_{2n} (\tau)$ are simple, and denoting its zeroes in the interior of $\Gamma$ by $\tau_i, i = 1, 2, \dots, n$, the
analyticity of $\psi_{1+}$ is obtained by imposing that the numerator of $\psi_{1+}$ in \eqref{psi12p} has zeroes of order at least 2 at $\tau_i$. Let us write
\bea
Q_{N_j} = \tau \, {\tilde Q}_{N_j -1} + A_j \;\;\;,\;\;\; j=1,2 ,
\label{QQN}
\eea
where the constants $A_1, A_2$  are determined by the normalization conditions 
$\psi_{1+}  (0) = 1, \psi_{2+}  (0) = 0$ for the $1^{\rm st}$ column of $G_+^{-1}$ and $\psi_{1+}  (0) = 0, \psi_{2+}  (0) = 1$ for the $2^{\rm nd}$ column.
By \eqref{ppsi} the
constants $A_1$ and $A_2$ cannot 
be simultaneously equal to $0$, see Remark \ref{remNNk}. The coefficients of ${\tilde Q}_{N_1 -1}, {\tilde Q}_{N_2 -1}$ will be given by the non-homogenous linear
system of $2n$ equations for the $2n$ unknown coefficients, 
\bea
&& \left[\tau \left(  {\tilde Q}_{N_1-1} \, \, \tau^{N_2 - k_{22}}  \,   {\tilde p}_{2 k_{22}} - {\tilde Q}_{N_2-1} \,  \tau^{N_1 - {k_{12}}} \, {\tilde p}_{2 k_{12}} 
 \right) \right] (\tau_i) = \nonumber\\
&& \qquad  - A_1 \, \, \tau_i^{N_2 - k_{22}}  \,   {\tilde p}_{2 k_{22}} (\tau_i) + A_2 \,  \tau_i^{N_1 - {k_{12}}} \, {\tilde p}_{2 k_{12}}  (\tau_i) , 
\nonumber\\
&& \left[\tau \left(  {\tilde Q}_{N_1-1} \, \, \tau^{N_2 - k_{22}}  \,   {\tilde p}_{2 k_{22}} - {\tilde Q}_{N_2-1} \, 
 \tau^{N_1 - {k_{12}}} \, {\tilde p}_{2 k_{12}}  \right) \right]' (\tau_i) = \nonumber\\
&& \qquad  - A_1 \,
 \left( \, \tau^{N_2 - k_{22}}  \,   {\tilde p}_{2 k_{22}} \right)' (\tau_i) + A_2 \,
\left(  \tau^{N_1 - {k_{12}}} \, {\tilde p}_{2 k_{12}} \right)' (\tau_i) , 
\eea
which is equivalent to 
\bea
\label{sysAA12}
&& \left[\left( {\tilde  Q}_{N_1-1} \,  \tau^{N_2 - k_{22}}  \,   {\tilde p}_{2 k_{22}} - {\tilde Q}_{N_2-1} \, 
 \tau^{N_1 - {k_{12}}} \, {\tilde p}_{2 k_{12}}   \right) \right] (\tau_i) = \nonumber\\
 && \qquad  - \frac{1}{\tau_i} \left( A_1 \, 
 \,  \tau^{N_2 - k_{22}}  \,   {\tilde p}_{2 k_{22}} (\tau_i) - A_2 \,  \tau_i^{N_1 - {k_{12}}} \, {\tilde p}_{2 k_{12}}  (\tau_i)\right)  , 
\\
&& \left[\left( {\tilde  Q}_{N_1-1} \,  \tau^{N_2 - k_{22}}  \,   {\tilde p}_{2 k_{22}} - {\tilde Q}_{N_2-1} \, 
 \tau^{N_1 - {k_{12}}} \, {\tilde p}_{2 k_{12}}   \right)
\right]' (\tau_i) = \nonumber\\
&& \qquad  - \frac{1}{\tau_i} \left[
A_1 \, \left( \,  \tau^{N_2 - k_{22}}  \,   {\tilde p}_{2 k_{22}} \right)' (\tau_i) -  A_2 \, \left(   \tau^{N_1 - {k_{12}}} \, {\tilde p}_{2 k_{12}} \right)'  (\tau_i) 
\right. \nonumber\\
&& \left. \qquad \qquad \qquad 
+ \frac{1}{\tau_i} \left( A_1 \, \tau_i^{N_2 - k_{22}}  \,   {\tilde p}_{2 k_{22}}  (\tau_i) 
- A_2 \,  \tau_i^{N_1 - {k_{12}}} \, {\tilde p}_{2 k_{12}}  (\tau_i)\right)  \right].
\nonumber
\eea

Finally, the fact the solutions are $C^{\infty}$ functions of $(\rho,v)$ can be verified directly.
\begin{flushright}  \qedsymbol{} \end{flushright}

\vskip 3mm

\noindent
{\it Proof of Theorem\ref{theoD0}:}\\
Note that the matrix coefficient of the system \eqref{sysAA12} is the same as that of \eqref{qqN}, so its determinant is given by $D(\rho, v)$ in Theorem \ref{theoNNn}.
On the other hand, by Cramer's rule, the coefficients of the polynomials $ {\tilde  Q}_{N_1-1}$ and  $ {\tilde  Q}_{N_2-1}$ are given by a quotient, where
the numerator is not zero (since $A_1$ and $A_2$ cannot vanish simultaneously) and the denominator is $D(\rho, v)$. Now, $1/\Delta$ is given by
the element in the second row and second column in $\displaystyle{\lim_{\tau \rightarrow \infty}} {\cal M}^-_{\rho, v} (\tau)$, i.e. according to 
\eqref{ppsi} and \eqref{QQN}, $1/\Delta$ is equal to the coefficient of the term of order $N_2 -1$ in ${\tilde Q}_{N_2-1} $, which tends to $\infty$
when $D(\rho,v) \rightarrow 0$. Since $g_{tt} = - \Delta$, we have
$g_{tt} \rightarrow 0$.

\begin{flushright}  \qedsymbol{} \end{flushright}

 \section{Examples in four and five space-time dimensions
 \label{sec:kerrMP}}

As an application, we consider in this section several examples which illustrate the results of Sections \ref{lsc} and \ref{sec:TOWH} as well as possible generalisations.

\subsection{The non-extremal Kerr monodromy matrix}

Let ${\cal M}$ be given by (see 
\cite{Woodhouse1997,Katsimpouri:2012ky})
\begin{equation}
{\cal M}(\omega)= \frac{1}{\omega^2 - c^2} \begin{pmatrix}
(\omega - m)^2 + a^2 & \quad 2 a m   \\
2 a m  & \quad (\omega + m)^2 + a^2 
\end{pmatrix} \;\;\;,\;\;\; c = \sqrt{m^2 - a^2} > 0 .
\label{Monokerr}
\end{equation}
By composition with $\omega_{\tau, \rho,v}$ we obtain a monodromy matrix of the form \eqref{calm22}, where 
\bea
\label{qMn2}
q_{2n}(\tau) &=& q_4 (\tau) = \frac14 \left[ \rho^2 (1 - \tau^2)^2 + 4 (v^2 - c^2) \tau^2 + 4  \rho v \tau (1-\tau^2) \right] \;\;\;\;\;\; (n=2) \;,
\\
\tilde{\cal M}_{\rho,v} (\tau) &=&  \begin{pmatrix}
 (v - m +  \rho \, \frac{(1- \tau^2)}{ 2 \tau } )^2 + a^2 &  2 a m  \\ 
 2 a m & (v + m +  \rho \, \frac{ (1- \tau^2)}{ 2 \tau} )^2 + a^2
\end{pmatrix} .
\eea
Note that this is a case considered in $(ii)$ in Theorem \ref{theoNNn}, with 
\bea
k_{11} = k_{22} = 2 \;\;\;,\;\;\; k_{12} = 0 \;\;\;,\;\;\;
N_1 = N_2 = 2 \;\;\;,\;\;\; 2n = 4 = N_1 + N_2 \;.
\eea
Let $\tau_1, {\tilde \tau}_1, \tau_2, {\tilde \tau}_2$ be the (simple) zeroes of $q_4$, in the notation of Section \ref{sec:TOWH}, and let $\Gamma$ be any admissible contour with $\tau_1, \tau_2$ in its interior.

The system \eqref{qqN} now takes the form of a system of 4 linear equations for the unknown coefficients $\alpha_0, \alpha_1, \beta_0, \beta_1$:
\bea
(\alpha_1 \tau_1 + \alpha_0 )  P_{22}(\tau_1) - 2 a m \tau_1^2 (  \beta_1 \tau_1 + \beta_0  ) &=& 0, \nonumber\\
\alpha_1  P_{22}(\tau_1) + (\alpha_1 \tau_1 + \alpha_0 )  P_{22}'(\tau_1)  - 4 a m \tau_1 (  \beta_1 \tau_1 + \beta_0  )
- 2 a m \tau_1^2  \beta_1 &=& 0, \nonumber\\
(\alpha_1 \tau_2 + \alpha_0 )  P_{22}(\tau_2) - 2 a m \tau_2^2 (  \beta_1 \tau_2 + \beta_0  ) &=& 0, \nonumber\\
\alpha_1   P_{22}(\tau_2) + (\alpha_1 \tau_2 + \alpha_0 )  P_{22}'(\tau_2)  - 4 a m \tau_2 (  \beta_1 \tau_2 + \beta_0  )
- 2 a m \tau_2^2  \beta_1 &=& 0,
\label{44sys}
\eea
which we can write as
\bea
 {\tilde T}  \begin{pmatrix}
 \alpha_0 \\ \alpha_{1} \\ \beta_0 \\  \beta_{1} 
 \end{pmatrix} = 0 ,
 \label{sysT2}
 \eea 
 where $\tilde T$ is the $4 \times 4$ matricial coefficient of the system \eqref{44sys}, with entries depending on the parameters $(\rho,v)$. The determinant $D(\rho,v)$ of  ${\tilde T}  $ is given by
 \bea
D(\rho,v) = f(\rho, v) \, h(\rho,v) ,
\label{Tfg}
\eea
where 
\bea
f(\rho,v) &=& \frac{a^2 m^2}{4} \, \rho^2 \left( \tau_1 - \tau_2 \right)^4 \neq 0 \;\;\;,\;\;\; \text{for all} \;\;  (\rho,v) \;,
\nonumber\\
h(\rho,v) &=& - 16  (m - v)^2 \tau_1^2 \tau_2^2 + \rho^2 \left( 1 + 4 \tau_1^3 \tau_2 + 6 \tau_1^2 \tau_2^2 + 4 \tau_1 \tau_2^3 + \tau_1^4 \tau_2^4 \right)\nonumber\\
&& - 8 \rho (m-v) \tau_1 \tau_2 \left( - \tau_1 - \tau_2 + \tau_1^2 \tau_2 + \tau_1 \tau_2^2 \right).
\label{funcfg}
\eea
Hence $D(\rho,v)  = 0$ if and only if 
\bea
h(\rho,v) = 0 \;.
\label{curg}
\eea
This condition describes a curve ${\cal C}$ in the Weyl half-plane
of the coordinates $\rho> 0, v \in \mathbb{R}$, which corresponds to the values of $(\rho,v)$ for which ${\cal M}_{\rho,v} (\tau)$ does not admit a canonical WH factorisation and for which we have 
$g_{tt} =0$, see Theorem \ref{theoD0}.
This curve ${\cal C}$ naturally depends on the choice of the points $\tau_1$ and $\tau_2$ that one chooses to be in the interior of $\Gamma$, which is the contour in the complex plane of the spectral parameter $\tau$ with respect to which the canonical factorisation of ${\cal M}_{\rho,v} (\tau)$ is sought.

If one looks for a canonical factorisation of ${\cal M}_{\rho,v} (\tau)$ w.r.t. an admissible contour $\Gamma$ such that 
\bea
\tau_1 &=&  \frac{ v-c - \sqrt{(v-c)^2 + \rho^2}}{\rho} , \nonumber\\
\tau_2  &=&   \frac{ v+c - \sqrt{(v+c)^2 + \rho^2}}{\rho} 
\eea
lie in $\mathbb{D}^+_{\Gamma}$, then that factorisation, obtained for $(\rho,v) \notin {\cal C}$, yields by \eqref{defM} a solution to the field equation \eqref{emotion} that describes the exterior region of the non-extremal Kerr black hole solution in General Relativity. The curve ${\cal C}$ defined by \eqref{curg} in the Weyl coordinates half-plane describes the ergosurface of that 
four-dimensional solution, as follows.
If we express \eqref{curg} in terms
of prolate spheroidal coordinates $(u,y)$ (see \cite{Harmark:2004rm,Katsimpouri:2012ky}),
\bea
v = u \, y \;\;\;,\;\;\; \rho= \sqrt{(u^2 - c^2)(1-y^2)},
\eea
where
\bea
c < u < + \infty \;\;\;,\;\;\; |y| < 1 \;,
\eea
we obtain
\bea
\rho \tau_1 &=& u y - c - (u - c y) , \nonumber\\
\rho \tau_2 &=& u y + c - (u + c y) .
\eea
Inserting these expressions into the expression of $h(\rho,v)$ given in \eqref{funcfg} we get
\bea
h(u,y) = \frac{16}{[(u^2 - m^2 + a^2)(1-y^2)]^3} (u^2 - c^2)^3 (y-1)^4 (u^2 - m^2 + a^2 y^2)
\eea
and we see that \eqref{curg} is equivalent to 
\bea
u (y) = \sqrt{m^2 - a^2 y^2} ,
\label{curuy}
\eea
This equation indeed defines 
the ergosurface of the non-extremal Kerr black hole in four dimensions, in the exterior region.
Namely, in standard Boyer-Lindquist coordinates, 
the metric component $g_{tt}$ of the Kerr solution is given by $g_{tt} = - ( r^2 - 2 m r + a^2 \cos^2 \theta)/( r^2 + a^2 \cos^2 \theta) $.
Converting from Boyer-Lindquist coordinates to coordinates $(u,y)$ using $u = r-m, \, y =  \cos \theta$ \cite{Harmark:2004rm}
shows that $g_{tt}$ vanishes precisely when \eqref{curuy} holds. The vanishing of $g_{tt}$ defines 
the ergosurface of the black hole.

\begin{figure}[hbt!]
	\centering
	\includegraphics[scale=1]{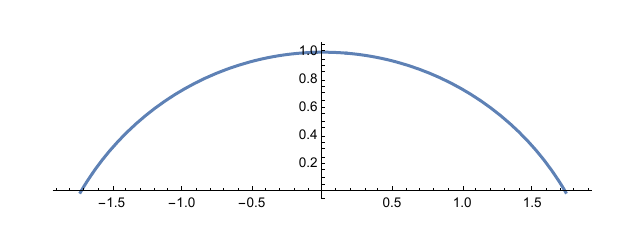} 
	\caption{Curve $\mathcal{C}$ in the Weyl coordinates upper half-plane $(\rho >0, v)$ for the values $m= 2, a = 1$. The horizontal axis represents $v \in \mathbb{R}$, while the vertical axis represents 
	$\rho> 0$.
	\label{ergos}}
\end{figure}

The curve $\cal C$ defined by $D(\rho,v) = 0$ 
is represented
 in Figure \ref{ergos}, 
  in the Weyl half-plane $\rho >0, v \in \mathbb{R}$.
The region between the curve $\mathcal{C}$  and the axis $\rho = 0$ 
represents the ergosphere, 
the region between the ergosurface $\mathcal{C}$  and the outer horizon of the
non-extremal Kerr black hole, while the complementary region describes the region outside of the ergosphere.

{\remark
Different choices of the contour $\Gamma$ yield different factorisations of ${\cal M}_{\rho, v} (\tau)$ and correspondingly different solutions $M(\rho,v)$. We can have 3 other cases: (i) ${\tilde \tau}_1$ and ${\tilde \tau}_2$ in $\mathbb{D}^+_{\Gamma}$, 
(ii) ${\tilde \tau}_1$ and ${\tau}_2$ in $\mathbb{D}^+_{\Gamma}$, 
(iii) ${\tau}_1$ and ${\tilde \tau}_2$ in $\mathbb{D}^+_{\Gamma}$.
In each case we obtain a curve $\cal C$ where 
the metric component $g_{tt}$ vanishes, $g_{tt} =0$; in case (i) it coincides with the curve for the ergosurface of the non-extremal Kerr black hole, while in the other two cases we obtain a different curve, namely
\bea
u(y) = \frac{c}{a} \sqrt{ m^2 - c^2 y^2} \;.
\eea
These various solutions reduce to solutions belonging to the class of A-metrics 
when the rotation parameter $a$ is set to $0$ (for the class of A-metrics see \cite{Aniceto:2019rhg} and references therein).}

\subsection{Rotating black hole in five space-time dimensions \label{sec:5drot}}

In this section we will discuss solutions to \eqref{emotion} with $\lambda = 1$ that yield solutions
in five space-time dimensions. Such solutions are described by a five-dimensional line element whose metric factors in Weyl coordinates are encoded in the $3 \times 3$ matrix $M(\rho,v)$ given in
\eqref{mrv5d}. In the notation used in \eqref{mrv5d}, the metric component $g_{tt}$ of the line element is given by 
$g_{tt} = - e^{2 \Sigma_3} + e^{2 \Sigma_2} \chi_1^2$, which is contained in the last component of the third column in \eqref{mrv5d}.

In the following we will consider the factorisation of two monodromy different matrices with respect to suitably chosen contours. When these matrices have a canonical factorisation, the resulting matrix factor \eqref{defM} has the form \eqref{mrv5d}. 
In the first example considered below,
the metric factor $g_{tt}$ is given by $g_{tt} =  - e^{2 \Sigma_3}$ since $\chi_1 =0$, whereas in the second example $g_{tt}$ takes the form $g_{tt} = - e^{2 \Sigma_3} + e^{2 \Sigma_2} \chi_1^2$.
In both cases, the resulting space-time solution describes the exterior region of a 
rotating Myers-Perry black hole solution in five space-time dimensions carrying one angular momentum $a$
\cite{Myers:1986un}, albeit in different space-time coordinates.
In one case, it yields the Myers-Perry solution in standard spherical coordinates \cite{Harmark:2004rm}, while in the other case it yields the Myers-Perry solution in the coordinates used in \cite{Chakrabarty:2014ora}.
Even though the functional form of $g_{tt}$ in both solutions looks different when expressed in terms
of the quantities $\Sigma_3$ and $\chi_1$, when converting to standard spherical coordinates the metric 
factor $g_{tt}$ takes the form \eqref{gtt5d} in both cases.

For each monodromy matrix, there is a curve $\mathcal C$ in the Weyl coordinate plane $\rho>0, v \in \mathbb{R}$ where the canonical factorisation does not exist. When approaching a point $(\rho,v)$ on the curve, the factor $e^{2 \Sigma_1}$ blows up, while the factor
$ e^{2 \Sigma_3}$ becomes vanishing. In the first example, $g_{tt} =  - e^{2 \Sigma_3}$ vanishes on the curve $\mathcal C$, which therefore coincides 
with the ergosurface in the exterior region of the black hole. In the second case, when approaching the corresponding curve $\mathcal C$, the combination $e^{2 \Sigma_2} \chi_1^2$ remains finite and $g_{tt}$
does not vanish on this curve. However, there are terms 
in $M(\rho,v)$, given in \eqref{mrv5d}, 
proportional to $e^{2 \Sigma_1}$,
that blow up on the curve $\mathcal C$.


First, let us then consider the following matrix (where $4 \alpha  = 2 m - a^2 > 0$), 
\begin{equation}
{\cal M}(\omega)= 
\begin{pmatrix}
\frac{\omega - \alpha + m}{2 (\omega + \alpha) (\omega - \alpha)} & 0 & \frac{ a m}{2 (\omega + \alpha) (\omega - \alpha)}
\\ 
  0 &  2(  \omega + \alpha) & 0 \\
\frac{ a m}{2 (\omega + \alpha) (\omega - \alpha)} & 0  & 
  \frac{ 8 (\omega - \alpha)^2 (\omega + \alpha) + 4 a^2 m^2}{8 (\omega + \alpha) (\omega - \alpha)(\omega - \alpha + m)}
\end{pmatrix} \;.
\label{MPmon}
\end{equation}
This matrix satisfies $\det {\cal M} = 1$ and has the property ${\cal M}^{\natural} (\omega) = {\cal M} (\omega)$, where ${\cal M}^{\natural} = \eta {\cal M}^T \eta$
with $\eta = {\rm diag} (1, -1, 1)$ \cite{Chakrabarty:2014ora}.

When the rotation parameter $a$ is set to zero, this reduces to the 
matrix
\begin{equation}
{\cal M}(\omega)= 
\begin{pmatrix}
\frac{1}{2 (\omega - \alpha)} & 0 & 0
\\ 
  0 &  2(  \omega + \alpha) & 0 \\
  0 & 0 & \frac{\omega - \alpha}{\omega + \alpha}
\end{pmatrix} 
\end{equation}
which, by composition with the spectral relation \eqref{spect} with $\lambda = 1$, yields a monodromy matrix 
whose canonical Wiener-Hopf factorisation with respect to a suitably chosen admissible contour $\Gamma$ gives the five-dimensional Schwarzschild black hole.
This contour is defined such that
\bea
\tau_{\alpha} &=&  \frac{ v-\alpha - \sqrt{(v-\alpha)^2 + \rho^2}}{\rho} , \nonumber\\
\tau_{- \alpha}  &=&   \frac{ v+\alpha - \sqrt{(v+\alpha)^2 + \rho^2}}{\rho} 
\label{tatmaMP}
\eea
lie in the interior of $\Gamma$. As as consequence,
${\tilde \tau}_{\alpha} = - 1/\tau_{\alpha}$ and ${\tilde \tau}_{-\alpha} = - 1/\tau_{-\alpha}$ lie in the exterior of $\Gamma$.

We compose \eqref{MPmon} with the spectral relation \eqref{spect} with $\lambda = 1$. The resulting monodromy matrix ${\cal M}_{\rho, v} (\tau)$ has poles in the $\tau$-plane located 
at $\tau_{\alpha}, \tau_{-\alpha}, {\tilde \tau}_{\alpha}, {\tilde \tau}_{-\alpha}$ and also at
\bea
\tau_{\alpha -m } =  \frac{ v-\alpha + m  - \sqrt{(v-\alpha + m)^2 + \rho^2}}{\rho}
\eea
and at ${\tilde \tau}_{\alpha -m } = - 1/ \tau_{\alpha -m } $. We will choose the contour $\Gamma$
with respect to which we will perform the canonical factorisation of ${\cal M}_{\rho, v} (\tau)$ to be such that the
{\it three} points $\tau_{\alpha}, \tau_{-\alpha}$ and $\tau_{\alpha -m}$ lie in its interior. The points
${\tilde \tau}_{\alpha}, {\tilde \tau}_{-\alpha}$ and ${\tilde \tau}_{\alpha -m}$ will then lie in the exterior of $\Gamma$.

Applying Theorem \ref{theo81} and studying the Riemann-Hilbert problem \eqref{rhgH} with $n=3$, we conclude that 
the canonical factorisation of ${\cal M}_{\rho, v} (\tau)$ with respect to $\Gamma$ exists for any point 
$(\rho >0,v)$
in the Weyl coordinates upper half-plane so long as $D(\rho, v) \neq 0$, where $D(\rho, v)$ is the determinant of a linear system that is
analogous to the one in \eqref{qqN} and is obtained by an entirely similar reasoning,
\bea
D(\rho, v) &=& \rho^6 \, \frac{\left( \tau_{\alpha} - \tau_{- \alpha} \right)^2 \left( \tau_{\alpha} - \tau_{ \alpha - m} \right)^2
\left( \tau_{- \alpha} - \tau_{ \alpha - m} \right)}{2 \tau^3_{\alpha} \, \tau_{-\alpha} \, \tau_{ \alpha - m}} 
( 1 + \tau_{\alpha} \tau_{ \alpha - m} )^3
\nonumber\\
&& \left[2 \tau_{\alpha} \tau_{ \alpha - m}  ( 1 + \tau_{\alpha} \tau_{ \alpha - m} ) ( 1 +  \tau_{- \alpha} \tau_{ \alpha - m})^2 - \tau_{-\alpha} \left( \tau_{\alpha} - \tau_{ \alpha - m} \right)^3 \, L
\right] ,
\label{D5d}
\eea
where $L$ denotes the ratio $L = a^2/m $ which, in view of the condition $4 \alpha  = 2 m - a^2 > 0 $, lies in the range $0 \leq L < 2$. Note that the bracket $( 1 + \tau_{\alpha} \tau_{ \alpha - m} )$ can be expressed 
as $(  {\tilde \tau}_{\alpha} -  \tau_{ \alpha - m} )/{\tilde \tau}_{\alpha}$. Hence it cannot vanish, since
${\tilde \tau}_{\alpha}$ and $\tau_{ \alpha - m}$ lie on different sides of the contour $\Gamma$.
Therefore, 
$D(\rho,v)$ can only vanish if the big bracket $[ \dots ]$ in \eqref{D5d} vanishes. To infer when this happens 
we express the coordinates $(\rho>0,v)$ in terms
of prolate spheroidal coordinates $(u,y)$ (see \cite{Harmark:2004rm,Chakrabarty:2014ora}),
\bea
v = \alpha \, u \, y \;\;\;,\;\;\; \rho= \alpha \sqrt{(u^2 - 1)(1-y^2)},
\label{prolH}
\eea
where
\bea
1 < u < + \infty \;\;\;,\;\;\; |y| < 1 \,.
\label{ranguy}
\eea
Then we have
\bea
\rho \tau_{\alpha} &=& \alpha \left( u y - 1 - (u -  y) \right) , \nonumber\\
\rho \tau_{-\alpha} &=& \alpha \left( u y + 1 - (u +  y) \right) , \nonumber\\
\rho \tau_{\alpha -m} &=& \alpha ( u y -1) + m - \sqrt{ (\alpha ( u y -1) + m )^2 + \alpha^2 (u^2 - 1)(1-y^2) } \;.
\eea
The big bracket $[ \dots ]$ in \eqref{D5d} becomes
\bea
2 \tau_{\alpha} \tau_{ \alpha - m}  ( 1 + \tau_{\alpha} \tau_{ \alpha - m} ) ( 1 +  \tau_{- \alpha} \tau_{ \alpha - m})^2 - \tau_{-\alpha} \left( \tau_{\alpha} - \tau_{ \alpha - m} \right)^3 \, L = \left( 2 + (L - 2) u - L y\right) \, \Sigma (u,y) \,, 
\label{bbSig}
\eea
with
\bea
\Sigma (u, y) = f_1 (u, y) + f_2 (u, y) \, \sqrt{f_3(u,y)} \,,
\eea
where $f_1, f_2, f_3$ are polynomials in $(u,y)$ given by
\bea
f_1(u, y) &=& - 4  \,  f_3(x,y)  \,  (y-1)^2 
\left( 4 (u-1) (y-1)^2 + L^2 (u-1) (y + 1)^2 - 4 L  ( 3 + u  + y^2 (u-1) \right) \;, \nonumber\\
f_2(u,y) &=& -4  (y-1 )^2 \left[ (L + 2) \left(16 L - (L-2)^2 u + (L-2)^2 u^2 \right) \right. \nonumber\\
&& \qquad \qquad \qquad + (L-2) (-1 + 
    2 u) \left(4 (-1 + u) + L^2 (-1 + u) 
     - 4 L (3 + u) \right) y \nonumber\\
     && \left. \qquad \qquad \qquad + (L-2)^2 (L + 2 ) (u-2) (u-1) y^2 - (L-2)^3 (u-1) y^3 \right]\;, 
\nonumber\\
f_3(u,y) &=& L^2 (u -y)^2 + 4 (u + y)^2 - 4 L (-2 + u^2 + y^2) \;.
\eea
It can be verified that $f_3$ and $\Sigma$ do not vanish when $(u,y)$ is the range \eqref{ranguy} with 
$0 \leq L < 2$. The combination \eqref{bbSig} can therefore only vanish on the line
\bea
 2 + (L - 2) u - L y = 0 \Longleftrightarrow y-1 + (u - y) \frac{2 \alpha}{m} = 0 \,.
 \label{ergoline}
 \eea
This line defines the ergosurface of the Myers-Perry black hole solution carrying angular momentum $a$,
where the metric component $g_{tt}$ vanishes. Therefore, we conclude that the canonical factorisation with respect
to the chosen contour $\Gamma$ exists for any point $(\rho,v)$
in the Weyl coordinates upper half-plane so long as it doesn't lie on the ergosurface of the rotating black hole.

By explicitly performing the canonical Wiener-Hopf factorisation \eqref{MMM} (when it exists) of 
${\cal M}_{\rho, v} (\tau)$
with respect to $\Gamma$ and using \eqref{defM},
we obtain for the solution $M(\rho, v)$, in the notation used in \cite{Chakrabarty:2014ora,Camara:2017hez}, 
\bea
M(\rho,v)= 
\begin{pmatrix}
e^{2 \Sigma_1} & 0 & e^{2 \Sigma_1} \chi_3 \\
0 & e^{2 \Sigma_2} & 0 \\
e^{2 \Sigma_1} \chi_3 & 0 & e^{2 \Sigma_3} + e^{2 \Sigma_1} \chi_3^2
\end{pmatrix} \;,
\label{matMP}
\eea
with
\bea
e^{2 \Sigma_2}&=& \sqrt{ \rho^2 + (v + \alpha)^2} + v + \alpha \;, \nonumber\\
e^{2 \Sigma_3} &=& \frac{ \sqrt{ \rho^2 + (v + \alpha)^2} + \sqrt{ \rho^2 + (v - \alpha)^2} 
\left( 1 - \frac{a^2}{m} \right) - 2 \alpha  }{ \sqrt{ \rho^2 + (v + \alpha)^2} + \sqrt{ \rho^2 + (v - \alpha)^2} 
\left( 1 - \frac{a^2}{m} \right) + 2 \alpha } \;, \nonumber\\
e^{2 \Sigma_1} &=& e^{- 2 \Sigma_2} e^{- 2 \Sigma_3} \;, \nonumber\\
\chi_3 &=& a  \frac{\sqrt{ \rho^2 + (v + \alpha)^2} - \sqrt{ \rho^2 + (v - \alpha)^2} + 2 \alpha}{ \sqrt{ \rho^2 + (v + \alpha)^2} + \sqrt{ \rho^2 + (v - \alpha)^2} 
\left( 1 - \frac{a^2}{m} \right) + 2 \alpha } \;.
\eea
This describes a Myers-Perry black hole solution, with one angular momentum $a$, in Weyl coordinates 
\cite{Harmark:2004rm}, which can be brought into standard form by converting 
the Weyl coordinates $(\rho,v)$ into standard spherical coordinates.
The $g_{tt}$ component of the space-time metric is given by 
$g_{tt} = - e^{2 \Sigma_3} $.

The entries of $M(\rho,v)$ in \eqref{matMP} have the following behaviour on the ergosurface:
$e^{2 \Sigma_3}$ vanishes, $e^{2 \Sigma_2}$ and $\chi_3$ are finite, while $e^{2 \Sigma_1}$ blows up.
Thus, various entries of $M(\rho,v)$ blow up when
approaching the ergosurface, similarly to what happens in the case of the Kerr black hole in four dimensions discussed above.

\vskip 3mm

Next, let us consider the matrix (where $4 \alpha  = 2 m - a^2 > 0$)
\bea
{\cal M}(\omega)= 
\begin{pmatrix}
- \frac{2}{\omega + \alpha} & 1 - \frac{m}{2 (\omega + \alpha)} & 0\\
- 1 + \frac{m}{2 (\omega + \alpha)} & - \frac{a^2 m}{4 (\omega - \alpha)} + \frac{m^2}{8(\omega + \alpha)} & \frac{a m}{2 (\omega - \alpha)} \\
0 & - \frac{a m}{2 (\omega - \alpha)} & 1 + \frac{m}{\omega - \alpha}
\end{pmatrix} \;.
\label{mvc}
\eea
This matrix is a special case of the one considered in 
\cite{Chakrabarty:2014ora}. The latter depends on two angular momenta denoted by $(l_1, l_2)$, while here we restrict ourselves
to one angular momentum, namely $a = l_1$.
This matrix satisfies $\det {\cal M} = 1$ and has the property ${\cal M}^{\natural} (\omega) = {\cal M} (\omega)$, where ${\cal M}^{\natural} = \eta {\cal M}^T \eta$
with $\eta = {\rm diag} (1, -1, 1)$ \cite{Chakrabarty:2014ora}. We now consider the composition of \eqref{mvc} with 
the spectral relation \eqref{spect} with $\lambda = 1$, and we denote the resulting monodromy matrix by 
${\cal M}_{\rho, v} (\tau)$.

When the rotation parameter $a$ is set to zero, we have $m = 2 \alpha$ and the matrix \eqref{mvc} reduces to the 
matrix
\begin{equation}
{\cal M}(\omega)= 
\begin{pmatrix}
- \frac{2}{\omega + \alpha} & \frac{\omega}{\omega + \alpha} & 0
\\ 
  - \frac{\omega}{\omega + \alpha} &  \frac{\alpha^2}{2(\omega + \alpha)}  & 0 \\
  0 & 0 & \frac{\omega + \alpha}{\omega - \alpha}
\end{pmatrix} 
\label{mon5dsch}
\end{equation}
which
results in a monodromy matrix 
${\cal M}_{\rho, v} (\tau)$,
whose canonical Wiener-Hopf factorisation with respect to a suitably chosen admissible contour $\Gamma$ gives a solution describing the exterior region of the five-dimensional Schwarzschild black hole.
This contour is the one that we choose in the following. It is defined such that
\bea
\tau_{\alpha} &=&  \frac{ v-\alpha + \sqrt{(v-\alpha)^2 + \rho^2}}{\rho} , \nonumber\\
\tau_{- \alpha}  &=&   \frac{ v+\alpha + \sqrt{(v+\alpha)^2 + \rho^2}}{\rho} 
\label{tatmacv}
\eea
lie in the interior of $\Gamma$. As as consequence,
${\tilde \tau}_{\alpha} = - 1/\tau_{\alpha}$ and ${\tilde \tau}_{-\alpha} = - 1/\tau_{-\alpha}$ lie in the exterior of $\Gamma$. Note that the choice of points \eqref{tatmacv} that lie in the interior of $\Gamma$ differs from the choice of points \eqref{tatmaMP} in the previous example. Therefore, the contour $\Gamma$ with respect to which
we perform the factorisation in this example differs from the one used in the previous example.

We now turn to the study of the existence of the canonical Wiener-Hopf factorisation of the monodromy
 ${\cal M}_{\rho, v} (\tau)$ associated with \eqref{mvc}. This is done as described in the proof of Theorem \ref{theoNNn}.
The expression for $\phi_{2 +}$ given in \eqref{cramphi12} now reads
\bea
\phi_{2+} = \frac{
\begin{vmatrix}
 \frac{4 \tau}{\rho ( \tau - {\tilde \tau}_{-\alpha}) }  \quad 
& k_1 \quad & 0 
\\
-1 \quad & 0 \quad & - \frac{a}{2} \\
0 \quad  & k_3 \quad & \tau - \tau_{\alpha} - \frac{2m \tau}{\rho(\tau - {\tilde \tau}_{\alpha})}
\end{vmatrix}
}{(\tau - \tau_{\alpha})(\tau - \tau_{-\alpha}) } = \frac{ k_1 \left( \tau - \tau_{\alpha} - \frac{2m \tau}{\rho(\tau - {\tilde \tau}_{\alpha}) }\right) + k_3 a \frac{2 \tau}{\rho(\tau - {\tilde \tau}_{-\alpha}) }}{(\tau - \tau_{\alpha})(\tau - \tau_{-\alpha}) } \;,
\eea
where $k_1, k_3 \in \mathbb{R}$, 
and analogously for $\phi_{1+}$. Imposing the analyticity of $\phi_{2+}$ at $\tau = \tau_{\alpha}$ and $\tau = \tau_{-\alpha}$ implies that
\bea
- k_1 \left( \frac{2m \tau_{\alpha}}{\rho(\tau_{\alpha} - {\tilde \tau}_{\alpha}) }\right) + k_3 a \frac{2 \tau_{\alpha}}{\rho(\tau_{\alpha} - {\tilde \tau}_{-\alpha})} &=& 0 \,, \nonumber\\
k_1 \left( \tau_{-\alpha}  - \tau_{\alpha} - \frac{2m \tau_{-\alpha}}{\rho(\tau_{-\alpha} - {\tilde \tau}_{\alpha}) }\right) + k_3 a \frac{2 \tau_{-\alpha}}{\rho(\tau_{-\alpha} - {\tilde \tau}_{-\alpha}) } &=& 0 \;.
\label{linsykk}
\eea
This is a linear system for the constants $k_1, k_3$. If the determinant of this linear system vanishes, the kernel of the associated Toeplitz operator has dimension one (c.f. the discussion around \eqref{rhgH}), and there is no canonical factorisation.
The determinant of the linear system \eqref{linsykk} reads
\bea
- \frac{2 a m \tau_{\alpha}}{\rho (\tau_{\alpha} - {\tilde \tau}_{\alpha})} \, D(\rho,v) \;,
\eea
with $D(\rho,v)$ given by
\bea
D(\rho,v) = \frac{\tau_{- \alpha}}{\rho}
\begin{vmatrix}
1 \quad & - \frac{1}{m} \frac{(\tau_{\alpha} - {\tilde \tau}_{\alpha})}{(\tau_{\alpha} - {\tilde \tau}_{- \alpha})} \\
- \frac{a^2}{(\tau_{-\alpha} - {\tilde \tau}_{\alpha})}  \quad  & \frac{2 }{(\tau_{-\alpha} - {\tilde \tau}_{-\alpha})}
\end{vmatrix}  \;.
\label{D5D}
\eea
Assuming $a \neq 0$, it follows that the vanishing of \eqref{D5D} occurs when $D(\rho,v) =0$, or equivalently
$D(\rho,v) = 0 \Leftrightarrow g(\rho,v) = 0$, where
\bea
g(\rho,v) =  2m (\tau_{\alpha} - {\tilde \tau}_{- \alpha}) (\tau_{-\alpha} - {\tilde \tau}_{\alpha}) - a^2 (\tau_{\alpha} - {\tilde \tau}_{\alpha}) (\tau_{-\alpha} - {\tilde \tau}_{-\alpha}) \;.
\eea
To study the condition $g(\rho,v) = 0$, we express the coordinates $(\rho>0,v)$ in terms
of the prolate spheroidal coordinates $(u,y)$ given in \eqref{prolH}.
Then, using 
\bea
\sqrt{( v \pm \alpha)^2 + \rho^2 } =  \alpha \left(  u \pm  y \right) > 0,
\eea
we obtain
\bea
\rho \tau_{\alpha} &=& \alpha \left( u y - 1 + u -  y \right) , \nonumber\\
\rho \tau_{-\alpha} &=& \alpha \left( u y + 1  + u +  y \right) ,
\eea
and the condition $g(\rho,v) = 0$ gives
\bea
u^2 - y^2 = \frac{m}{2 \alpha} (1 - y^2)\;.
\label{condu2v2}
\eea
This defines a curve $\mathcal{C}$ in the Weyl coordinates upper half-plane $(\rho >0, v)$.
On this curve, $e^{2 \Sigma_3}$ vanishes and 
the quantities $\chi_1, \chi_2, \chi_3$ and $e^{2 \Sigma_2}$ remain finite, but 
the matrix $M(\rho,v)$ is ill behaved, since its matrix entry \eqref{sigm1blow}
blows up. We note, however, 
that the metric component $g_{tt}$, given in \eqref{gtt5d},
stays finite and non-zero on $\mathcal{C}$: although the 
condition \eqref{condu2v2}  resembles the condition \eqref{ergoline} for the ergosurface of the rotating Myers-Perry black hole with one angular momentum $a$, the former does not equal the latter.

We have explicitly performed the canonical Wiener-Hopf factorisation \eqref{MMM} (assuming its existence) of 
${\cal M}_{\rho, v} (\tau)$
with respect to $\Gamma$. The resulting expression for $M(\rho, v)$ is given in \eqref{5dvcfact}. The solution
$M(\rho, v)$ describes the exterior region of 
a rotating Myers-Perry black hole solution with one angular momentum $a$, as can be verified
by converting 
the Weyl coordinates $(\rho,v)$ into the coordinates used in \cite{Chakrabarty:2014ora}.

\vskip 3mm

Summarising, in the examples discussed above we have shown that, as in the $2 \times 2$ case,
 the non-existence of a canonical Wiener-Hopf factorisation of the
 monodromy matrices occurs on simple smooth curves 
 in the Weyl coordinates upper-half plane.
 However, unlike in the the $2 \times 2$ case considered in Section \ref{sec:TOWH} (see Theorem \ref{theoD0}), 
these 
curves may or may not correspond to ergosurfaces, which are defined by 
the norm of the Killing vector $\partial /\partial t$ becoming null there.

\vskip 2mm

\subsection*{Acknowledgements}
The authors would like to thank the Isaac Newton Institute for Mathematical Sciences, Cambridge, for support and hospitality during the programme
{\it Black holes: bridges between number theory and holographic quantum information},
where part of the work on this paper was undertaken. This work was supported by EPSRC grant no EP/R014604/1.
This work was partially
supported by FCT/Portugal through CAMGSD, IST-ID, projects UIDB/04459/2020 and UIDP/04459/2020.


\appendix

\section{Canonical factorisation of a $3 \times 3$ monodromy matrix \label{sec:A}}

The canonical factorisation \eqref{MMM} (if it exists) of a monodromy matrix ${\cal M}_{\rho,v} (\tau)$ with respect
to an admissible contour $\Gamma$ is performed as in \eqref{gpsps}. Here we consider the monodromy matrix given in \cite{Chakrabarty:2014ora}, obtained by 
composition of \eqref{mvc} with 
the spectral relation \eqref{spect} with $\lambda = 1$. Its canonical factorisation with respect to the contour
$\Gamma$ specified below \eqref{mon5dsch} results in a matrix \eqref{defM} given by
\bea
M(\rho,v) = 
\begin{pmatrix}
A_{11} & A_{12} & A_{13} \\
- 1 - \frac{m}{4} A_{11} + \frac{a}{2} A_{31} \quad & \frac{m}{4}  - \frac{m}{4} A_{12} + \frac{a}{2} A_{32} \quad &
- \frac{a}{2}  - \frac{m}{4} A_{13} + \frac{a}{2} A_{33} \\
A_{31} & A_{32} & A_{33} 
\end{pmatrix}\;,
\label{5dvcfact}
\eea
where 
\bea
A_{11} &=& \frac{4 \tau_{- \alpha} \left( \tau_{\alpha} - {\tilde \tau}_{\alpha} \right)}{m \rho^2 \left( \tau_{- \alpha} - {\tilde \tau}_{- \alpha} \right) \left( \tau_{\alpha} - {\tilde \tau}_{- \alpha} \right) D}
\left( \frac{2m}{\tau_{\alpha} - {\tilde \tau}_{\alpha}} - \frac{a^2}{\tau_{- \alpha} - {\tilde \tau}_{\alpha} }
\right) \nonumber\\
A_{12} &=&  1 + \frac{m}{4} A_{11} - \frac{a}{2} A_{31} \nonumber\\
A_{31} &=& A_{13} = - \frac{4 a \tau_{- \alpha} }{\rho^2 \left( \tau_{- \alpha} - {\tilde \tau}_{ \alpha} \right)
D} \left( \frac{1}{\tau_{- \alpha} - {\tilde \tau}_{- \alpha}} - \frac{1}{\tau_{\alpha} - {\tilde \tau}_{-\alpha} }
\right) \\
A_{32} &=& \frac{a}{2}  + \frac{m}{4} A_{13} - \frac{a}{2} A_{33} \nonumber\\
A_{33} &=& \frac{1}{m \rho D} 
\left[\frac{2m}{ \left( \tau_{- \alpha} - {\tilde \tau}_{ -\alpha} \right) }
\left( \tau_{\alpha} - \frac{a^2 \tau_{-\alpha}}{\rho ( \tau_{- \alpha} - {\tilde \tau}_{ \alpha})} \right) -
\frac{a^2 \tau_{- \alpha} }{ ( \tau_{- \alpha} - {\tilde \tau}_{ \alpha} ) ( \tau_{\alpha} - {\tilde \tau}_{ -\alpha} )} \left( \tau_{\alpha} - {\tilde \tau}_{\alpha} - \frac{2m }{\rho} \right)
\right] \;,\nonumber
\eea
with $D$ given by \eqref{D5D}. 

The matrix $M(\rho,v)$ is of the form \cite{Chakrabarty:2014ora,Camara:2017hez}
\bea
\begin{pmatrix}
e^{2 \Sigma_1}& e^{2 \Sigma_1} \chi_2 & e^{2 \Sigma_1} \chi_3 \\
- e^{2 \Sigma_1} \chi_2 \quad &   -  e^{2 \Sigma_1} \chi_2^2 + e^{2 \Sigma_2}\quad &
 -  e^{2 \Sigma_1} \chi_2 \chi_3 + e^{2 \Sigma_2}\chi_1\\
e^{2 \Sigma_1} \chi_3 &  e^{2 \Sigma_1} \chi_2 \chi_3 - e^{2 \Sigma_2}\chi_1  & 
 -  e^{2 \Sigma_2} \chi_1^2 +  e^{2 \Sigma_1} \chi_3^2  + e^{2 \Sigma_3} 
 \label{mrv5d}
\end{pmatrix}
\eea
with $\Sigma_1 + \Sigma_2 + \Sigma_3 = 0$ 
and satisfies $M^{\natural}= M$, where $M^{\natural} = \eta M^T \eta$ with $\eta = {\rm diag} (1,-1,1)$,
as required for a coset representative $M \in SL(3,\mathbb{R})/SO(2,1)$. 
The five-dimensional space-time metric
is then expressed in terms of these matrix entries as (c.f. eqs. (A.1)-(A.6) in \cite{Chakrabarty:2014ora})
\bea
ds^2_5 = e^{2 \Sigma_1} \, ds_3^2 - e^{2 \Sigma_3} \left( dt + {\cal A}_2 \right)^2 + e^{2 \Sigma_2} \left( d \psi + \chi_1 \, dt + {\cal A}_1 \right)^2 \,
\eea
with
\bea
ds_3^2 = f^2 \left( d \rho^2 + dv^2 \right) + \rho^2 \, d \phi^2 \;.
\eea
The scalars $\Sigma_1, \Sigma_2, \Sigma_3, \chi_1, f$ and 
the 1-forms ${\cal A}_1$ and ${\cal A}_2$ are functions of $\rho, v$.
The two 1-forms are dualised into scalars $\chi_2$ and $\chi_3$ 
using
\bea
e^{-(4 \Sigma_1 + 2 \Sigma_3)} \, \star_3  {\cal F}_1 = d \chi_2 
\;\;\;,\;\;\; - e^{- 2 (\Sigma_1 - \Sigma_3)} \, \star_3  {\cal F}_2 = d \chi_3 - \chi_1 d \chi_2 \;,
\eea
where
\bea
{\cal F}_1 = d {\cal A}_1 + {\cal A}_2 \wedge d \chi_1 \;\;\;,\;\;\;
{\cal F}_2 = d {\cal A}_2 \;,
\eea
and where $ \star_3$ denotes the Hodge star operator in three dimensions.  Finally, the function $f^2$ is determined from $M(\rho,v)$ by integration, as in eq. (2.7) in \cite{Aniceto:2019rhg}.

In particular,
the metric component $g_{tt}$ is given by 
\bea
- g_{tt} = e^{2 \Sigma_3} - e^{2 \Sigma_2} \chi_1^2 = A_{33} - \frac{A_{13}^2}{A_{11}} = \frac{ u - y - \frac{m}{2 \alpha} (1 -y)}{ u - y + \frac{m}{2 \alpha} (1 +y ) } = 1 - \frac{2m}{r^2 + a^2 \cos^2 \theta} \;,
\label{gtt5d}
\eea
where we converted from prolate spheroidal coordinates $(u,y)$ to spherical coordinates $(r, \theta)$ using the relations
 $r^2 = 2 \alpha ( u + 1), \, 2 \cos^2 \theta = y +1$ 
\cite{Harmark:2004rm,Katsimpouri:2012ky}.

When approaching a point on the curve $\mathcal{C}$ specified by \eqref{D5D} in a non-tangential manner, the matrix entry $e^{2 \Sigma_1}$, 
\bea
e^{2 \Sigma_1} = \frac{2}{\alpha} \left[  \frac{  u - y + \frac{m}{2 \alpha} (1 + y) }{  u^2 - y^2 - \frac{m}{
2 \alpha} (1 -y^2) } \right] \;,
\label{sigm1blow}
\eea
blows up, while  $g_{tt}$ remains finite. This metric component vanishes when $u - y = \frac{m}{2 \alpha} (1 -y)$, which defines the ergosurface of the rotating Myers-Perry black hole.


\begingroup\raggedright

\endgroup


\begin{thebibliography}{99}




 \bibitem{AAM}
V.~M. Adukov, N.~V. Adukova, and G.~Mishuris, {\it An explicit Wiener-Hopf
  factorisation algorithm for matrix polynomials and its exact realizations
  within ExactMPF package},  {\em Proceedings of the Royal Society A:
  Mathematical, Physical and Engineering Sciences} {\bf 478} (2022), no.~2263
  20210941,
\href{https://doi.org/10.1098/rspa.2021.0941}{{\tt
  https://doi.org/10.1098/rspa.2021.0941}}.


\bibitem{Aniceto:2019rhg}
P.~Aniceto, M.~C. C\^amara, G.~L. Cardoso, and M.~Rossell\'o, {\it {Weyl
  metrics and Wiener-Hopf factorisation}},  {\em JHEP} {\bf 05} (2020) 124,
  \href{https://arxiv.org/pdf/1910.10632.pdf}{{\tt 1910.10632}}.


\bibitem{Belinsky:1971nt}
V.~A. Belinsky and V.~E. Zakharov, {\it {Integration of the Einstein Equations
  by the Inverse Scattering Problem Technique and the Calculation of the Exact
  Soliton Solutions}},  {\em Sov. Phys. JETP} {\bf 48} (1978) 985--994.




\bibitem{Breitenlohner:1986um}
P.~Breitenlohner and D.~Maison, {\it {On the Geroch Group}},  {\em Ann. Inst.
  H. Poincare Phys. Theor.} {\bf 46} (1987) 215.

\bibitem{Breitenlohner:1987dg}
P.~Breitenlohner, D.~Maison, and G.~W. Gibbons, {\it {Four-Dimensional Black Holes from Kaluza-Klein Theories}},  {\em Commun. Math. Phys.} {\bf 120} (1988) 295.

\bibitem{Camara:2022gvc}
M.~C.~C\^amara and G.~L.~Cardoso, \emph{{Generating new gravitational solutions by matrix multiplication}},  
{\em Proceedings of the Royal Society A:
  Mathematical, Physical and Engineering Sciences} {\bf 480} (2024), no.~2285,
\href{https://doi.org/10.1098/rspa.2023.0857}{{\tt
  https://doi.org/10.1098/rspa.2023.0857}},
\href{https://arxiv.org/pdf/2211.01702.pdf}{{\tt 2211.01702}}.


\bibitem{Camara:2017hez}
M.~C. C\^amara, G.~L. Cardoso, T.~Mohaupt, and S.~Nampuri, {\it {A
  Riemann-Hilbert approach to rotating attractors}},  {\em JHEP} {\bf 06}
  (2017) 123, \href{https://arxiv.org/pdf/1703.10366.pdf}{{\tt 1703.10366}}.

\bibitem{CDR}  
 M.~C. C\^amara, C. Diogo, and L. Rodman,
  {\it {Fredholmness of Toeplitz operators and corona problems}}, {\em J. Funct. Anal.} {\bf  259}  (2010) 1273-1299. 


\bibitem{COP}
M.~C. C\^amara, R. O'Loughlin, and J.~R. Partington, {\it Dual-band general Toeplitz operators},
{\em 
 Mediterr. J. Math.} { \bf  19}  (2022) 175,
 \href{https://arxiv.org/pdf/2012.14725.pdf}{{\tt 2012.14725}}.
 
 

\bibitem{CP}
M.~C. C\^amara and J.~R. Partington,
{\it Spectral properties of truncated Toeplitz operators by equivalence after extension}, 
{\em J. Math. Anal. Appl.} {\bf  433}  (2016) 762-784.


\bibitem{Cardoso:2017cgi}
G.~L. Cardoso and J.~C. Serra, {\it {New gravitational solutions via a
  Riemann-Hilbert approach}},  {\em JHEP} {\bf 03} (2018) 080,
  \href{https://arxiv.org/pdf/1711.01113.pdf}{{\tt 1711.01113}}.
  
    \bibitem{Chakrabarty:2014ora}
B.~Chakrabarty and A.~Virmani, {\it {Geroch Group Description of Black Holes}},
   {\em JHEP} {\bf 11} (2014) 068,
  [\href{https://arxiv.org/pdf/1408.0875.pdf}{{\tt 1408.0875}}].



  \bibitem{CG} K.~F. Clancey and I.~Gohberg, {\it {Factorisation of Matrix Functions and Singular Integral Operators}},
in {\em {Operator Theory: Advances and Applications, vol. 3, Birkh\"auser Verlag, Basel, 1981}}.


  
 \bibitem{De}
 P. Deift, {\it Riemann-Hilbert problems}, in 
 {\em Random matrices, pp. 1-40, IAS/Park City Math. Ser. 26, Amer. Math. Soc., Providence, RI, 2019},
  \href{https://arxiv.org/pdf/1903.08304.pdf}{{\tt 1903.08304}}.

 
 
\bibitem{Du}
R. Duduchava, {\it {Integral Equations with Fixed Singularities}},  {\em
  Teubner, Leipzig, 1979.}

\bibitem{Figueras:2009mc}
P.~Figueras, E.~Jamsin, J.V.~Rocha and A.~Virmani, \emph{{Integrability of Five Dimensional Minimal Supergravity and Charged Rotating Black Holes}}, \href{https://doi.org/10.1088/0264-9381/27/13/135011}{\emph{Class. Quant. Grav.} {\bfseries 27} (2010) 135011} [\href{https://arxiv.org/pdf/0912.3199.pdf}{{\tt 0912.3199}}].



\bibitem{Gohberg2003FactorizationAI}
I.~Gohberg, N.~Manojlovic, and A.~F. dos Santos, {\it Factorization and
  integrable systems : summer school in Faro, Portugal, September 2000},  2003.


\bibitem{GKR}
G.~J. {Groenewald}, M.~A. {Kaashoek}, and A.~C.~M. {Ran}, {\it {Wiener-Hopf
  factorisation indices of rational matrix functions with respect to the unit
  circle in terms of realization}},  {\em  Indag. Math.} {\bf 34} (2023), no. 2, 338 - 356,
   \href{https://arxiv.org/pdf/2203.07821.pdf}{{\tt
  2203.07821}}.

\bibitem{Harmark:2004rm}
T.~Harmark, \emph{{Stationary and axisymmetric solutions of higher-dimensional general relativity}}, \href{https://doi.org/10.1103/PhysRevD.70.124002}{\emph{Phys. Rev. D} {\bfseries 70} (2004) 124002} 
[\href{https://arxiv.org/pdf/hep-th/0408141.pdf}{{\tt hep-th/0408141}}].



\bibitem{Its}
A.~R. Its, {\it {The Riemann-Hilbert Problem and Integrable Systems}},  {\em
  Notices of the AMS} {\bf 50} (2003) 1389.
  
 \bibitem{Katsimpouri:2012ky}
D.~Katsimpouri, A.~Kleinschmidt, and A.~Virmani, {\it {Inverse Scattering and
  the Geroch Group}},  {\em JHEP} {\bf 02} (2013) 011,
  \href{https://arxiv.org/pdf/1211.3044.pdf}{{\tt 1211.3044}}.

\bibitem{KisilAMR}
A.~V. Kisil, I.~D. Abrahams, G.~Mishuris, and S.~V. Rogosin, {\it The
  Wiener-Hopf technique, its generalizations and applications: constructive
  and approximate methods},  {\em Proceedings of the Royal Society A:
  Mathematical, Physical and Engineering Sciences} {\bf 477} (2021), no.~2254
  20210533,
  \href{https://doi.org/10.1098/rspa.2021.0533}{{\tt
  https://doi.org/10.1098/rspa.2021.0533}}.



\bibitem{KA}
V.~D. {Kunz} and R.~C. {Assier}, {\it {Diffraction by a Right-Angled
  No-Contrast Penetrable Wedge Revisited: A Double Wiener-Hopf Approach}},
  {\em SIAM J. Appl. Math.} {\bf 82} (2022), no. 4, 1495 - 1519,
  \href{https://arxiv.org/pdf/2112.03173.pdf}{{\tt 2112.03173}}.



\bibitem{LS}
G.~Litvinchuk and I.~Spitkovsky, {\it {factorisation of Measurable Matrix
  Functions}},  in {\em {Oper. Theory Adv. Appl., vol. 25, Birkh\"auser Verlag,
  Basel, 1987. Translated from Russian by B. Luderer, with a foreword by B.
  Silbermann}}.


\bibitem{Lu:2007jc}
H.~Lu, M.~J. Perry, and C.~N. Pope, {\it {Infinite-dimensional symmetries of
  two-dimensional coset models coupled to gravity}},  {\em Nucl. Phys. B} {\bf
  806} (2009) 656--683, \href{https://arxiv.org/pdf/0712.0615.pdf}{{\tt
  0712.0615}}.


\bibitem{MS}
E. Meister, E. and F.-O. Speck, {\it Modern Wiener-Hopf methods in diffraction theory}, in {\em Ordinary and partial differential equations, 
Vol. II, Proc. 10th Dundee Conf., Dundee/UK 1988. Pitman Res. Notes Math. Ser. 216, 130-171 (1989).}



\bibitem{MP}
S.~Mikhlin and S.~Pr\"ossdorf, {\it {Singular integral operators}},   {\em
  {Springer-Verlag, Berlin, 1986. Translated from German by Albrecht B\"ottcher
  and Reinhard Lehmann}}.

\bibitem{Myers:1986un}
R.C.~Myers and M.J.~Perry, \emph{{Black Holes in Higher Dimensional Space-Times}}, \href{https://doi.org/10.1016/0003-4916(86)90186-7}{\emph{Annals Phys.} {\bfseries 172} (1986) 304}.



\bibitem{Nicolai:1991tt}
H.~Nicolai, {\it {Two-dimensional gravities and supergravities as integrable
  system}},  {\em Lect. Notes Phys.} {\bf 396} (1991) 231--273.



\bibitem{Penna:2021kua}
R.~F. Penna, {\it {Einstein\textendash{}Rosen waves and the Geroch group}},
  {\em J. Math. Phys.} {\bf 62} (2021), no.~8 082503,
  \href{https://arxiv.org/pdf/2106.13252.pdf}{{\tt 2106.13252}}.




\bibitem{Schwarz:1995af}
J.~H. Schwarz, {\it {Classical symmetries of some two-dimensional models
  coupled to gravity}},  {\em Nucl. Phys. B} {\bf 454} (1995) 427--448,
  \href{https://arxiv.org/pdf/hep-th/9506076.pdf}{{\tt hep-th/9506076}}.



\bibitem{Woodhouse1997}
N.~Woodhouse, {\it Integrability and Einstein's equations},  {\em Banach Center
  Publications} {\bf 41} (1997), no.~1 221--232.



\end{thebibliography}
\end{document}